\RequirePackage{lineno} 
\documentclass[aps,prc,superscriptaddress,showpacs,floatfix,nofootinbib,twocolumn]{revtex4}
\usepackage{graphicx} 
\usepackage{float} 
\usepackage{amssymb}
\usepackage{url}
\usepackage{harpoon}
\usepackage[colorlinks,breaklinks]{hyperref}
\hypersetup{linkcolor=blue,citecolor=blue,filecolor=black,urlcolor=blue}
\newcommand{\pT} {\ensuremath{p_{\mathrm{T}}}}

\begin{document}
\title{Disentangling flow and nonflow correlations via Bayesian unfolding of the event-by-event distributions of harmonic coefficients in ultrarelativistic heavy-ion collisions}
\newcommand{\sunysb}{Department of Chemistry, Stony Brook University, Stony Brook, NY 11794, USA}
\newcommand{\bnl}{Physics Department, Brookhaven National Laboratory, Upton, NY 11796, USA}
\author{Jiangyong Jia}\email[Correspond to\ ]{jjia@bnl.gov}
\affiliation{\sunysb}\affiliation{\bnl}
\author{Soumya Mohapatra}\affiliation{\sunysb}
\begin{abstract}
The performance of the Bayesian unfolding method in extracting the event-by-event (EbyE) distributions of harmonic flow coefficients $v_n$ is investigated using a toy model simulation, as well as simulations based on the HIJING and AMPT models. The unfolding method is shown to recover the input $v_2-v_4$ distributions for events with multiplicities similar to those observed in Pb+Pb collisions at the LHC.  The effects of the nonflow are evaluated using HIJING simulation with and without a flow afterburner. The probability distribution of $v_n$ from nonflow is nearly a Gaussian and can be largely suppressed with the data-driven unfolding method used by the ATLAS Collaboration. The residual nonflow effects have no appreciable impact on the $v_3$ distributions, but affect the tails of the $v_2$ and $v_4$ distributions; these effects manifest as a small simultaneous change in the mean and standard deviation of the $v_n$ distributions. For the AMPT model, which contains both flow fluctuations and nonflow effects, the reduced shape of the extracted $v_n$ distributions is found to be independent of $\pT$ in the low $\pT$ region, similar to what is observed in the ATLAS data. The prospect of using the EbyE distribution of the harmonic spectrum aided by the unfolding technique as a general tool to study azimuthal correlations in high energy collisions is also discussed.
\end{abstract}
\pacs{25.75.Dw} \maketitle 

\section{Introduction}
In recent years, the measurement of harmonic flow coefficients $v_n$ has provided important insight into the hot and dense matter created in heavy ion collisions at the Relativistic Heavy Ion Collider (RHIC) and the Large Hadron Collider (LHC). These coefficients are generally obtained from a Fourier expansion of the particle distribution in azimuthal angle $\phi$:
\begin{equation}
\label{eq:flow}
\frac{dN}{d\phi}\propto1+2\sum_{n=1}^{\infty}v_{n}\cos n(\phi-\Phi_{n})\;,
\end{equation}
where $v_n$ and $\Phi_n$ represent the magnitude and phase (event plane or EP angle) of the $n^{\mathrm{th}}$-order anisotropy. The coefficients $v_n$ stem from the hydrodynamic response to various shape components or eccentricities $\epsilon_n$ of the created matter. In non-central heavy ion collisions, $\epsilon_2$ is mainly associated with the ``elliptic'' shape of the nuclear overlap region. However, $\epsilon_2$ in central collisions and the other $\epsilon_n$ terms arise primarily from fluctuations of the nucleon positions in the overlap region~\cite{Alver:2010gr}. Model calculations suggest that $v_n$ scales nearly linearly with $\epsilon_n$, for $n=2$ and 3~\cite{Qiu:2011iv}. The proportionality constant is found to be sensitive to the properties of the matter, such as the equation of state and shear viscosity~\cite{Voloshin:2008dg,Teaney:2009qa}. 

Measurements of $v_n$ coefficients for $n=1$--6~\cite{Adare:2011tg,star:2013wf,Aamodt:2011by,Aad:2012bu,CMS:2012wg} have established the importance of the fluctuations in the initial state, and provided quantitative constraints on the transport properties of the created matter. Most of these measurements focus on the $v_n$ values averaged over many events, which mainly reflect the hydrodynamic response of the created matter to the average collision geometry in the initial state. Estimation of the second moment of $v_n$ distribution for n=2 has been obtained via a Monte Carlo template fit~\cite{Alver:2007qw}, or two- and four-particle cumulant methods~\cite{Agakishiev:2011eq,Abelev:2012di}. Recently the ATLAS Collaboration made the first measurement of the probability distribution of charged hadron $v_n$, $p(v_n)$~\cite{Jia:2012ve}, as well as the correlations between two or three EP angles of different order~\cite{Jia:2012sa} in Pb+Pb collisions at $\sqrt{s_{NN}}=2.76$ TeV. These new types of flow measurements provide additional information on the nature of the fluctuations in the initial geometry and the non-linear effects in the hydrodynamic evolution\cite{Teaney:2012ke,Qiu:2012uy,Teaney:2012gu,Gale:2012rq,Niemi:2012aj}.

%are calculated using charged particles in the pseudorapidity range $|\eta|<2.5$ and the transverse momentum range $\pT>0.5$ GeV, which are then
The study of these new observables requires new experimental techniques. ATLAS extracts the $p(v_n)$ using a data-driven Bayesian unfolding method based on the event-by-event (EbyE) distribution of either the single-particle $\phi$ or the pair $\Delta\phi$~\cite{Jia:2012ve}. The key step in the ATLAS measurement is to determine a response function which connects the observed signal $v_n^{{\mathrm{obs}}}$ with the true $v_n$ signal, $p(v_n^{{\mathrm{obs}}}|v_n)$. The response function is then used to unfold the $v_n^{{\mathrm{obs}}}$ distributions to obtain the true $v_n$ distributions, using the Bayesian unfolding technique~\cite{Agostini,unfold}. The response function mainly corrects for the smearing due to the effect of finite per-event charged particle multiplicity, but is also shown to reduce nonflow effects. The efficacy of the method has been demonstrated using many data-driven cross-checks, however, a detailed Monte-Carlo study can provide a direct check of the performance of the method, as well as a measure of the influence of nonflow effects.

In this paper, the performance of Bayesian unfolding, used by the ATLAS Collaboration to extract the $v_n$ distribution, is studied using the HIJING~\cite{Gyulassy:1994ew} and AMPT~\cite{Lin:2004en} event generators. We show that the unfolding procedure indeed accounts for the effects associated with finite multiplicity and nonflow. The structure of the paper is as follows: Section~\ref{sec:1} gives a brief review of the unfolding method developed for the data analysis. Section~\ref{sec:2} demonstrates the basic performance of the unfolding procedure via a toy model simulation that includes only statistical smearing. The influence of nonflow effects is studied in Section~\ref{sec:3} by applying the unfolding procedure to HIJING events with and without a flow afterburner. Section~\ref{sec:4} gives a study of the $v_n$ distributions in AMPT events. Section~\ref{sec:5} gives a summary of the results. Section~\ref{sec:6} discusses the prospect of applying the method to the study of the azimuthal correlation in high-energy collisions.

\begin{figure*}[t!]
\centering
\includegraphics[width=0.8\linewidth]{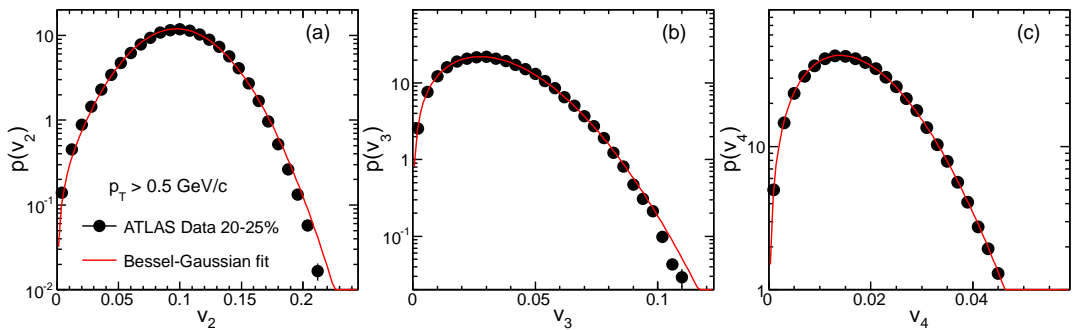}
\caption{\label{fig:in} (Color online) The ATLAS $v_n$ data~\cite{Jia:2012ve} and parameterization by Bessel-Gaussian functions. These functions are used as the input distribution for $v_n$ in the toy model and HIJING simulations.}
\end{figure*}

\section{Unfolding procedure}
\label{sec:1}
The ATLAS analysis method starts with a Fourier expansion of the azimuthal distribution of charged particles in a given event:
\begin{eqnarray}
\nonumber
\frac{dN}{d\phi}&\propto& 1+2\sum_{n=1}^{\infty}v_n^{\mathrm{obs}}\cos n(\phi-\Phi_{n}^{\mathrm{obs}})\\\nonumber
&=& 1+2\sum_{n=1}^{\infty}\left(v_{n,\mathrm{x}}^{\mathrm{obs}}\cos n\phi+v_{n,\mathrm{y}}^{\mathrm{obs}}\sin n\phi)\right),\\\nonumber
v_n^{\mathrm{obs}}&=& \sqrt{\left(v_{n,\mathrm{x}}^{\mathrm{obs}}\right)^2+\left(v_{n,\mathrm{y}}^{\mathrm{obs}}\right)^2},\mathrm{tan}\;n\Phi_{n}^{\mathrm{obs}}=\frac{v_{n,\mathrm{y}}^{\mathrm{obs}}}{v_{n,\mathrm{x}}^{\mathrm{obs}}}\\\label{eq:1}
\end{eqnarray}
where $v_n^{\mathrm{obs}}$ is the magnitude of the observed per-particle flow vector: $\overrightharp{v}_n^{\;\mathrm{obs}}=(v_{n,\mathrm{x}}^{\mathrm{obs}},v_{n,\mathrm{y}}^{\mathrm{obs}})$. In the limit of large multiplicity and the absence of nonflow effects, it approaches the true flow signal: $v_n^{\mathrm{obs}}\rightarrow v_n$. Hence one needs to determine the response function $p(\overrightharp{v}_n^{\;\mathrm{obs}}|\overrightharp{v}_n)$ or $p(v_n^{\mathrm{obs}}|v_n)$, which can be used to unfold these smearing effects.

The response function is determined via a two-subevents method (2SE): the charged particles are divided into two equal subevents with $\eta>0$ and $\eta<0$. The smearing effects are estimated from the distribution of the difference of the flow vectors between the two subevents, $p_{\mathrm{sub}}\left((\overrightharp{v}_n^{\;\mathrm{obs}})^a-(\overrightharp{v}_n^{\;\mathrm{obs}})^b\right)$, for which the physical flow signal cancels. This distribution is observed to be well described by a 2D Gaussian with identical widths, $\delta_{_{\mathrm{2SE}}}$, in both dimensions and hence the response function can be obtained as:
\begin{eqnarray}
\label{eq:2a}
p(\overrightharp{v}_n^{\;\mathrm{obs}}|\overrightharp{v}_n)&=&\frac{1}{2}p_{\mathrm{sub}}\left(2\left[(\overrightharp{v}_n^{\;\mathrm{obs}})^a-(\overrightharp{v}_n^{\;\mathrm{obs}})^b\right]\right) \\\label{eq:2b}
&\approx&\frac{1}{2\pi\delta_n^2} e^{-\frac{\left(\overrightharp{v}^{\mathrm{\;obs}}_n-\overrightharp{v}_n\right)^2}{2\delta_n^2}}, \delta_n=\delta_{_{\mathrm{2SE}}}/2.
\end{eqnarray}
where the factor of two accounts for the fact that the smearing in $\overrightharp{v}_n^{\;\mathrm{obs}}$ is a factor of two narrower than that for $(\overrightharp{v}_n^{\;\mathrm{obs}})^a-(\overrightharp{v}_n^{\;\mathrm{obs}})^b$. Consequently the 1D response function can be obtained by integrating out the azimuthal angle:
\begin{eqnarray}
\label{eq:3}
p(v_n^{\mathrm{obs}}|v_n) \approx\frac{1}{2\pi\delta_n^2} v_n^{\mathrm{obs}}e^{-\frac{(v_n^{\mathrm{obs}})^2+v_n^2}{2\delta_n^2}} I_0\left(\frac{v_n^{\mathrm{obs}}v_n}{\delta_n^2}\right).
\end{eqnarray}
where $I_0$ is the modified Bessel function of the first kind. Eq.~\ref{eq:3} is known as the Bessel-Gaussian (BG) function~\cite{Ollitrault:1992bk,Voloshin:2007pc}, which is a very good approximation of the response function. In ATLAS data analysis, the actual distribution, obtained by shifting the measured 2D distribution Eq.~\ref{eq:2a} to $\overrightharp{v}_n$ and integrating out the azimuthal angle, is used in the unfolding. 

The Bayesian unfolding procedure from~\cite{Agostini} is used to obtain the $v_n$ distribution.  In this procedure, the true $v_n$ (cause ``c'') distribution is obtained from the measured $v_n^{\mathrm{obs}}$ (effect ``e'') distribution and the response function $A_{ji}\equiv p(e_j|c_i)$, using an iterative procedure:
\begin{eqnarray}
\label{eq:bay1}
\hat{c}^{\mathrm{iter}+1} = \hat{M}^{\mathrm{iter}}\hat{e}, \;\;\; M_{ij}^{\mathrm{iter}} = \frac{A_{ji}c_i^{\mathrm{iter}}}{\sum_{m,k}A_{mi}A_{jk}c_k^{\mathrm{iter}}}\;.
\end{eqnarray}
The unfolding matrix $\hat{M}^{0}$ is determined from the response function and initial guess of the true distribution $\hat{c}^0$ (referred to as prior), and the process is iterated. The $v_2$ and $v_3$ distributions were shown to converge after a few iterations, while more iterations ($\geq 16$) were needed for $v_4$. One of the goals of this paper is to verify the convergence of the unfolding via simulations.

The ATLAS results also show that the response function (Eq.~\ref{eq:3}) not only accounts for the smearing effect, but also for the nonflow contributions. The argument goes as follows: most of the nonflow effects arise from localized clusters each containing a few particles, and the number of such clusters is proportional to the total multiplicity. Therefore, the nonflow contributions to $v_n^{\;\mathrm{obs}}=|\overrightharp{v}_n^{\;\mathrm{obs}}|$ is of a similar nature as that for the response function which is obtained from $(\overrightharp{v}_n^{\;\mathrm{obs}})^a-(\overrightharp{v}_n^{\;\mathrm{obs}})^b$. Thus, most of the short-range nonflow effects can be removed by the unfolding procedure. This can be verified directly via a Monte-Carlo event generator such as HIJING~\cite{Gyulassy:1994ew} as discussed later.

\begin{figure*}[t!]
\centering
\includegraphics[width=0.8\linewidth]{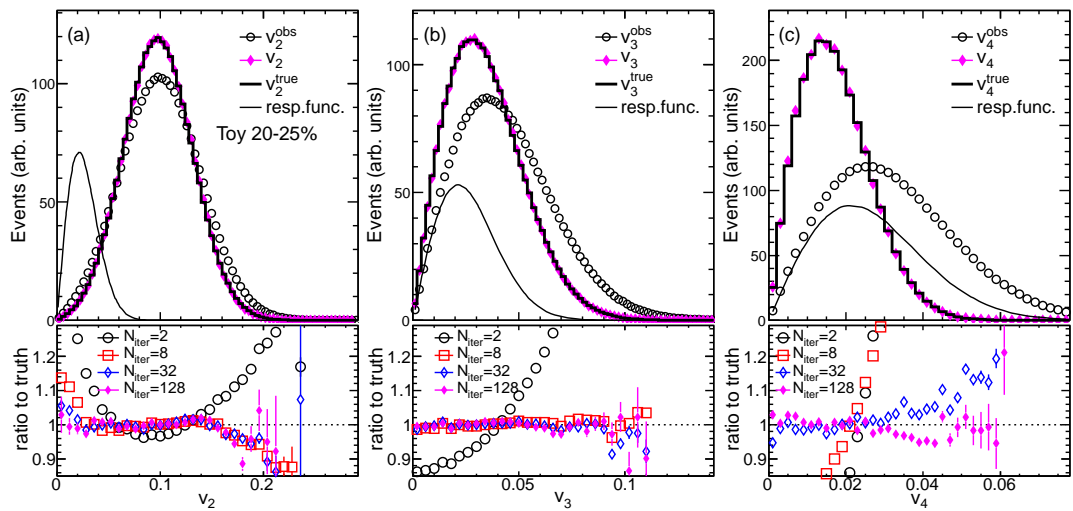}
\caption{\label{fig:toy} (Color online) The performance of the Bayesian unfolding for $v_2$ (left), $v_3$ (middle) and $v_4$  (right) in toy simulation.}
\end{figure*}

\section{Toy model simulation}
\label{sec:2}
The toy model simulation serves as a proof-of-principle check for the unfolding method. The multiplicity of each event is generated according to a Gaussian distribution with a mean of $\langle N\rangle=1070$ and a width of $\sigma_N =84$. This roughly simulates the distribution of the number of reconstructed charged particles for $\pT>0.5$ GeV/$c$ and $|\eta|<2.5$ in 20-25\% Pb+Pb collisions at $\sqrt{s_{NN}}=2.76$ TeV in ATLAS, assuming a 70\% reconstruction efficiency~\cite{trk}. The input $v_n$ signal is obtained by a Bessel-Gaussian function that is parameterized to the ATLAS data as shown in Figure~\ref{fig:in}. Particles in a given event are implemented to have the same $v_n$ value, which is sampled from the BG functions. The particles are then randomly divided into two subevents with equal multiplicity. The correlations of the per-particle flow vectors between the two subevents are then used to calculate the response function. About 2.5 Million events were generated; this is similar to the statistics used in the ATLAS data analysis for the 20-25\% centrality bin.

Figure~\ref{fig:toy} shows the performance of the Bayesian unfolding for different number of iteration ($N_{\mathrm{iter}}$). The response function (thin solid curves) are significantly narrower than the $v_{n}^{\mathrm{obs}}$ distribution for $n=2$ and $n=3$, leading to convergence to the input distribution after a few iterations (see ratios for $N_{\mathrm{iter}}=8$). However the convergence for $n=4$ is slower due to its much smaller $v_n$ signal. This overall performance is similar to what is observed in the data analysis~\cite{Jia:2012ve}.

\section{HIJING simulation}
\label{sec:3}
Understanding the nonflow effects in any flow measurement is a topic of intense debate over the last several years~\cite{Voloshin:2008dg,Ollitrault:2009ie,Jia:2006sb,Alver:2010rt,Agakishiev:2011eq,Abelev:2012di}. In this paper, the nonflow correlations refer to statistical smearing due to a finite number of observed particles, short-range correlations among few particles such as resonance decays, Bose-Einstein correlation and fragments of individual jets, as well as long-range correlations not related to flow such as Global momentum conservation (GMC) and fragments from di-jets. Note that most flow methods are not sensitive to pure statistical smearing, but it is listed as part of nonflow since it appears explicitly in the EbyE distributions. Also note that the GMC effect is a global correlation that mainly affects the $v_1$, and the correlations between two back-to-back jets involve a few particles but can be long-range in $\eta$, affecting mainly the even harmonics. These nonflow correlations and their fluctuations are naturally included in the HIJING event generator and has been used frequently in previous investigations of nonflow effects~\cite{Jia:2006sb,Alver:2010rt,Kikola:2011tu,Xu:2012ue}.

Since HIJING does not have physical flow, any non-zero $v_n^{\mathrm{obs}}$ values must be solely due to nonflow. However, we can also impose a physical flow signal on the particles using a flow afterburner, such that the generated events contain both the nonflow and flow contributions that fluctuate event-by-event. The Bayesian unfolding procedure can be used to extract the $v_n$ distributions for comparison with the input. This allows a quantitative estimate of how the nonflow effects and their EbyE fluctuations affect the final extracted $v_n$ distributions.

\begin{figure}[b]
\centering
\includegraphics[width=1.0\columnwidth]{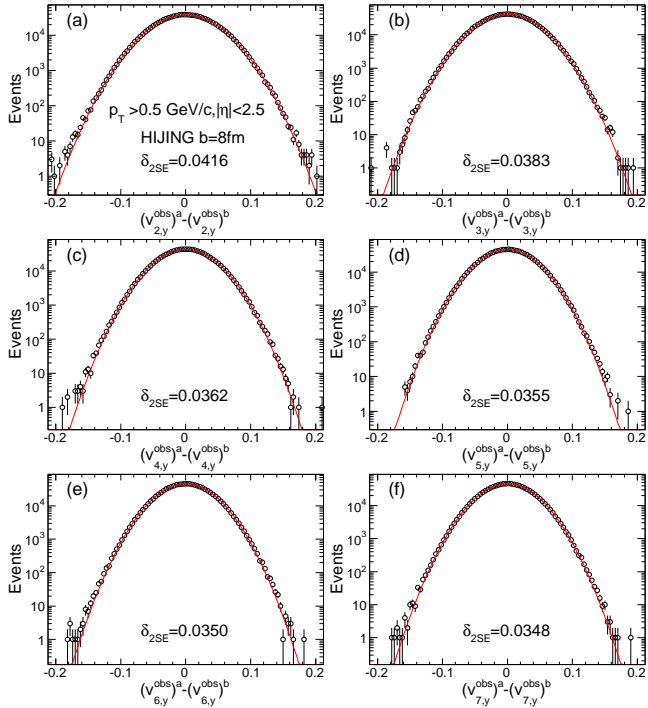}
\caption{\label{fig:hia0} (Color online) The distribution of $(\protect\overrightharp{v}_n^{\;\mathrm{obs}})^a-(\protect\overrightharp{v}_n^{\;\mathrm{obs}})^b$ projected to the $y$-axis for HIJING events at $b=8$~fm. Each panel shows the distribution for one harmonic number. Lines indicate a fit to a Gaussian function.}
\end{figure}
\begin{figure}[b]
\centering
\includegraphics[width=1.0\columnwidth]{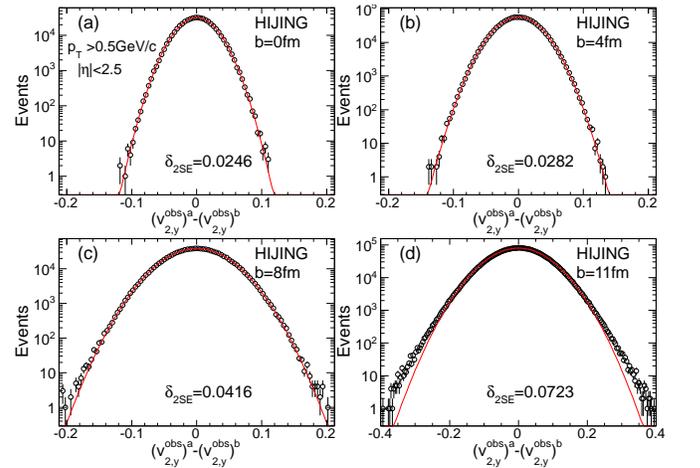}
\caption{\label{fig:hia0b} (Color online) The distribution of $(\protect\overrightharp{v}_n^{\;\mathrm{obs}})^a-(\protect\overrightharp{v}_n^{\;\mathrm{obs}})^b$ projected to the $y$-axis for n=2 from HIJING events. Each panel shows the distribution for one impact parameter. Lines indicate a fit to a Gaussian function.}
\end{figure}

\begin{figure*}[!t]
\centering
\includegraphics[width=1.0\linewidth]{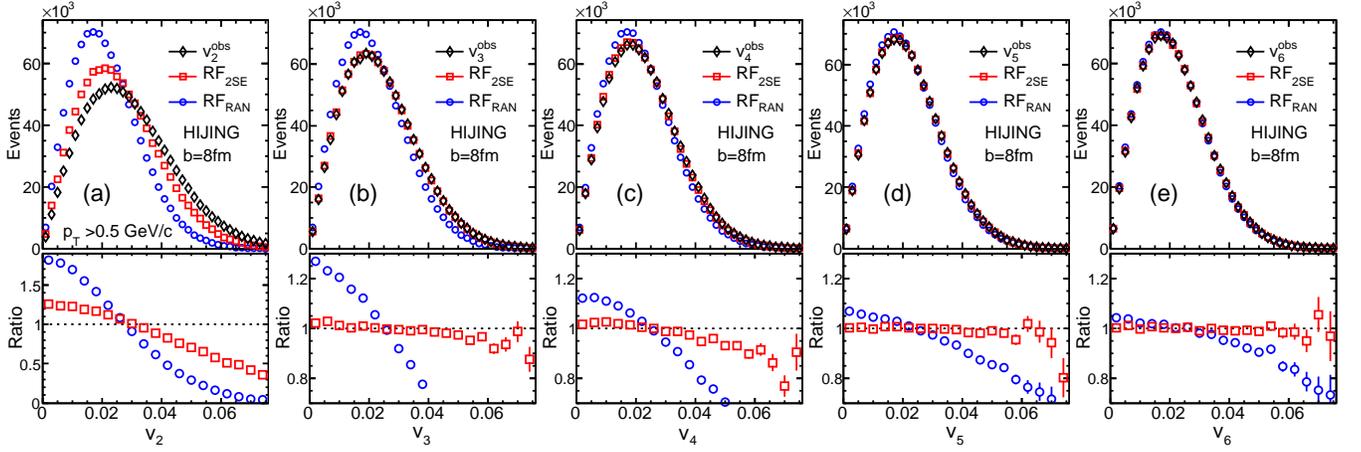}
\caption{\label{fig:hia1} (Color online) Top panels: distributions of $v_n^{\;\mathrm{obs}}$ (diamond), response function obtained via standard 2SE method (squares) and response function obtained via two subevents with randomization of tracks in $\phi$ (circles). Bottom panels: The ratios of two response functions to the $v_n^{\;\mathrm{obs}}$ distribution. Each column shows the result for one harmonic number ($n=2$--6 from left to right).}
\end{figure*}
In this study, we generated 0.5--3 million HIJING events each at four fixed impact parameters, $b=0$~fm, $b=4$~fm, $b=8$~fm and $b=11$~fm for Pb+Pb collisions at 2.76 TeV. The event multiplicity roughly corresponds to 0-1\%, 5-10\%, 20-25\% and 50-55\% centrality ranges in the data. The experimentally measured $v_n$ distributions in these four centrality ranges are parameterized individually by the BG functions. These functions are then used as the input $v_n$ distributions to be imposed on the HIJING events. The unfolding procedure is repeated for all four impact parameters, but the results for $b=8$~fm are often used for illustrative purposes. In order to gain maximum sensitivity to the nonflow effects,  all particles, including neutral particles, above 0.5 GeV/$c$ and $|\eta|<2.5$ are used.

\subsection{Default HIJING (without flow)}
\label{sec:3.1}

The goal of HIJING simulation without a flow afterburner is to understand the basic properties of the response function and the contributions of non-flow effects. Figure~\ref{fig:hia0} shows the $y$-axis projections (the $x$-axis projections are identical) of the difference of the per-particle flow vector between the two symmetric subevents ($\eta>0$ and $\eta<0$), $p_{\mathrm{sub}}\left((\overrightharp{v}_n^{\;\mathrm{obs}})^a-(\overrightharp{v}_n^{\;\mathrm{obs}})^b\right)$. They are shown for $n=2$--7 in events with $b=8$~fm. These distributions are well described by a Gaussian, implying that the smearing effects and those due to nonflow short-range correlations are largely statistical. Figure~\ref{fig:hia0b} shows similar distributions for various impact parameters for $n=2$. Significant non-Gaussian behavior is observed for peripheral collisions (i.e. $b=11$~fm). In this case, the distribution can be described by the Student's t-distribution~\cite{Jia:2012ve}, which is a general probability distribution function for the difference between two estimates of the mean from independent samples. The t-distribution approaches a Gaussian distribution when the event multiplicity is large. The distributions shown in Figures~\ref{fig:hia0} and \ref{fig:hia0b} are used to obtain the response functions according to Eq.~\ref{eq:2a}.

Since $\overrightharp{v}_n=0$, in the limit of pure statistical smearing, the response function reduces to the $v_n^{\;\mathrm{obs}}$ distribution, and is a 2D Gaussian centered around the origin:
\begin{eqnarray}
\label{eq:hia0}
p(\overrightharp{v}_n^{\;\mathrm{obs}}) &=& p(\overrightharp{v}_n^{\;\mathrm{obs}}|\overrightharp{v}_n=0)= \frac{1}{2\pi\delta^2_{n}} e^{-\left(\overrightharp{v}_n^{\mathrm{obs}}\right)^2/\left(2\delta^2_{n}\right)}\\\label{eq:hia0b}
p(v_n^{\;\mathrm{obs}}) &=& \frac{v_n^{\;\mathrm{obs}}}{\delta^2_{n}} e^{-\left(v_n^{\mathrm{obs}}\right)^2/\left(2\delta^2_{n}\right)}
\end{eqnarray}
The $x$ or $y$ width of the 2D response function or $\delta_n$ can be calculated from the central limit theorem as (see for example Ref.~\cite{Ollitrault:1992bk,Ollitrault:2009ie}):
\begin{eqnarray}
\label{eq:hia1}
\delta_n=\frac{1}{N}\sqrt{\sum_{\nu=1}^{N}w_{\nu}^2\cos^2n\phi_{\nu}}=\sqrt{\frac{1}{2N}\frac{\langle w^2\rangle}{\langle w\rangle^2}} = \sqrt{\frac{1}{2N}}
\end{eqnarray}
where $N$ is total number of particles used in the event and the weight for each particle $w$ is taken to be one ($w=1$). This width is the limit expected for pure statistical smearing. Additional nonflow correlations tend to increase this width. For example, if the $N$ particles come from $N_r$ resonance decays each with $M$ number of particles with the same $\phi$ angle, the distribution is still a Gaussian but with a larger width:
\begin{eqnarray}
\label{eq:hia2}
\delta_n'= \sqrt{\frac{1}{2N_r}}= \sqrt{\frac{M}{2N}}=\sqrt{M}\delta_n
\end{eqnarray}

In the data analysis, the response function is obtained by 2SE method via Eq.~\ref{eq:hia1}, which has a slightly broader width:
\begin{eqnarray}
\nonumber
\delta_n^{''}= \frac{1}{2}\sqrt{\frac{1}{2N_a}+\frac{1}{2N_b}} = \sqrt{\frac{1}{2N(1-\alpha^2)}} \approx \delta_n (1+0.5\alpha^2)\\\label{eq:hia1a}
\end{eqnarray}
where $\alpha=(N_a-N_b)/(N_a+N_b)$ accounts for the fact that the two subevents may not have the same multiplicity event-by-event. If the fluctuations between the two subevents are assumed to be random, one expect $\alpha^2\approx 1/N$. The average $\alpha^2$ values are calculated directly in HIJING for the four impact parameters, and they are found to be $\langle\alpha^2\rangle=$0.05\%, 0.07\%, 0.14\% and 0.4\%, respectively.  This implies that the relative multiplicity fluctuation between the subevents has negligible effects as long as the input $v_n$ distribution and total multiplicity $N=N_a+N_b$ remain unchanged.

Eq.~\ref{eq:hia1} is valid for events with fixed multiplicity. The spread of the particle multiplicity in the event class can further increase the width:
\begin{eqnarray}
\label{eq:hia1b}
\delta_n=\sqrt{\left\langle\frac{1}{2N}\right\rangle}\approx \sqrt{\frac{1}{2\langle N\rangle}\left[1+\left(\frac{\sigma_N}{\langle N\rangle}\right)^2\right]}
\end{eqnarray}
where $\sigma_N$ is the standard deviation of the particle multiplicity distribution.

\begin{figure}[!h]
\centering
\includegraphics[width=1.0\columnwidth]{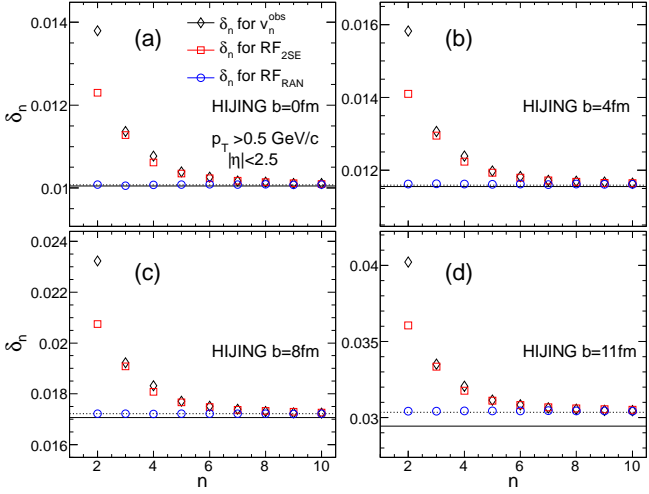}
\caption{\label{fig:hia2} (Color online) The widths of the distributions vs $n$ for $v_n^{\;\mathrm{obs}}$, RF$_{\mathrm{2SE}}$ and RF$_{\mathrm{RAN}}$ in four centrality classes in HIJING without flow. The solid lines indicate the values calculated via Eq.~\ref{eq:hia1} using the average multiplicity, while the dashed lines indicates the values taking into account the multiplicity fluctuation Eq.~\ref{eq:hia1b}. These distributions can be regarded as the power spectrum for nonflow in HIJING.}
\end{figure}

The argument in Eq.~\ref{eq:hia2} can be generalized to other short-range correlations as long as the number of sources are proportional to the total multiplicity. On the other hand, long-range correlations such as di-jets can correlate between the two subevents (with each subevent containing one jet), and hence their influence to the response function may differ from Eq.~\ref{eq:hia2}. Figure~\ref{fig:hia0} shows that the distributions of $(\overrightharp{v}_n^{\;\mathrm{obs}})^a-(\overrightharp{v}_n^{\;\mathrm{obs}})^b$ from HIJING are consistent with a Gaussian, but their widths decrease with harmonic number $n$. The width for $n=7$ is very close to the pure statistical fluctuation limit (see Figure~\ref{fig:hia2}). This behavior suggests that the nonflow contributions are largely statistical in nature, and their influence to $v_n$ decrease with the harmonic number.

Figure~\ref{fig:hia1} compares the distribution of $v_n^{\;\mathrm{obs}}$ and the response function obtained using the standard 2SE method (denoted as RF$_{\mathrm{2SE}}$), as well as the response function obtained by randomly dividing the particles into equal halves (denoted as RF$_{\mathrm{RAN}}$). Since each particle is randomly assigned to one of the subevent, any source of azimuthal correlation (include flow and non-flow) contributes equally to the two subevents, and hence cancels out in the distribution of $(\overrightharp{v}_n^{\;\mathrm{obs}})^a-(\overrightharp{v}_n^{\;\mathrm{obs}})^b$. Thereby the resulting response function is expected to have a width consistent with Eq.~\ref{eq:hia1b}, which is exactly what is found in HIJING simulation as shown by Figure~\ref{fig:hia1} (see also Figure~\ref{fig:hia2}). 

If RF$_{\mathrm{2SE}}$ properly accounts for all nonflow effects, it should coincide with the $v_n^{\;\mathrm{obs}}$ distribution. Figure~\ref{fig:hia1} shows that this is true for most harmonics, except for $n=2$ and to a lesser extent for $n=4$. These differences could be due to residual effects of di-jets, which affect mainly the even harmonics. It is also interesting to point out that the three distributions approach each other for large $n$ and reach the statistical limit given by Eq.~\ref{eq:hia1b} for $n>6$. This behavior is quite natural, as $n^{\mathrm{th}}$-order harmonics are sensitive to correlations at an angular scale of $2\pi/n$. The harmonics at very large $n$ eventually resolve the $\phi$ angle of all particles individually, and hence depend only on the total multiplicity in the event.

\begin{figure}[t]
\centering
\includegraphics[width=0.9\columnwidth]{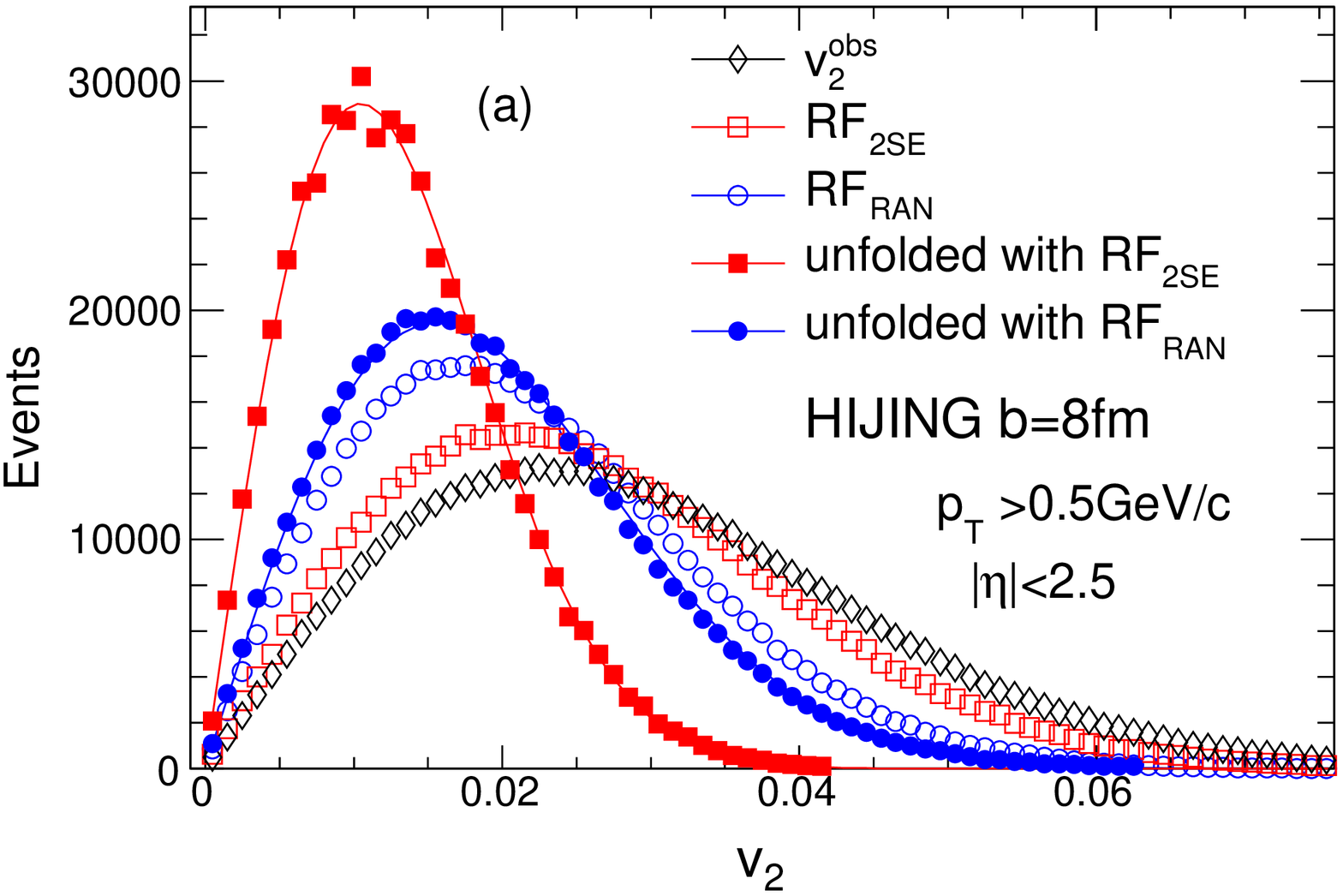}
\includegraphics[width=0.9\columnwidth]{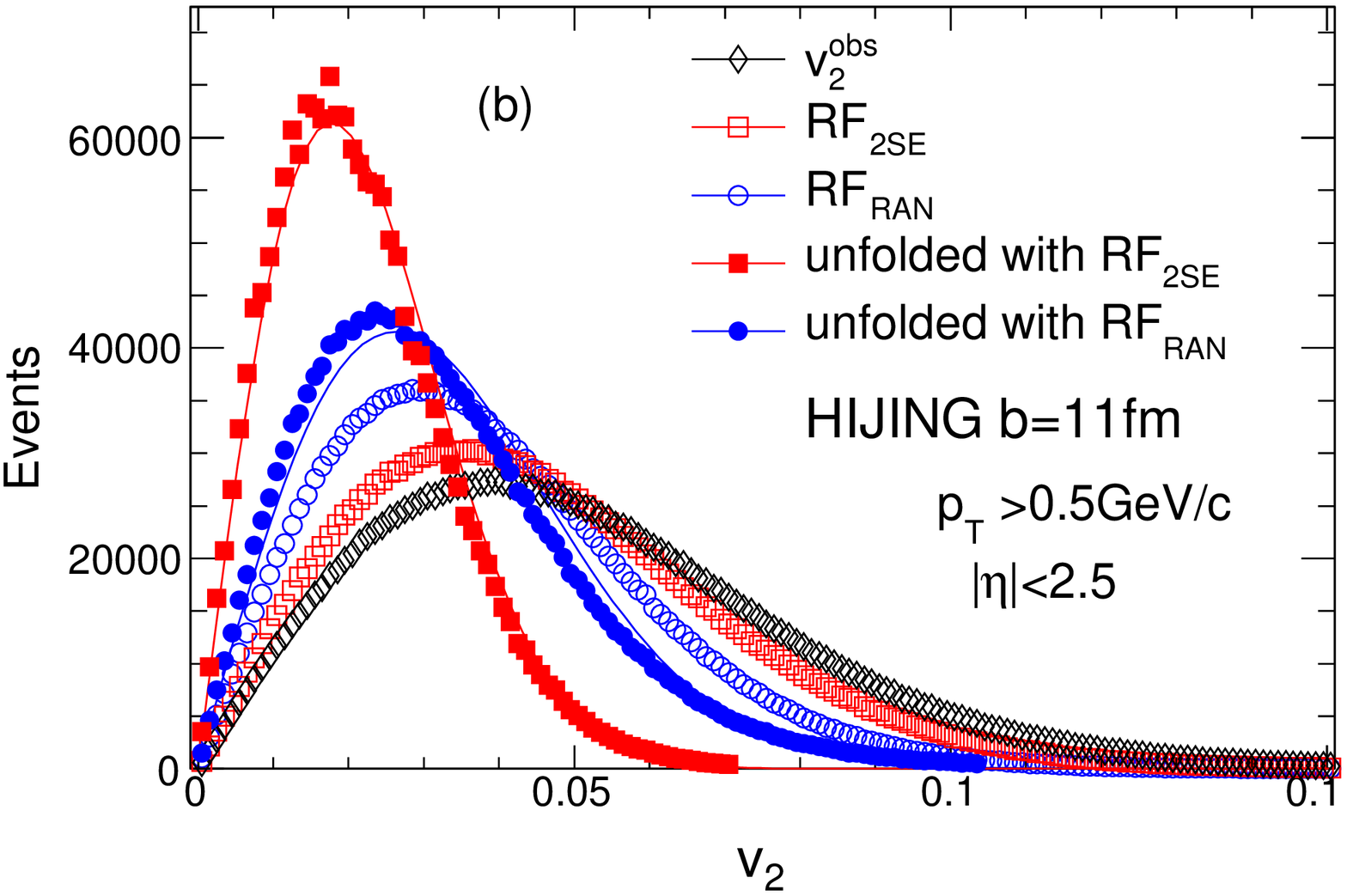}
\caption{\label{fig:hia3} (Color online) The distributions of $v_2^{\;\mathrm{obs}}$ (open diamonds) and the two response functions RF$_{\mathrm{2SE}}$ (open squares) and  RF$_{\mathrm{RAN}}$ (open circles), together with the estimated residual nonflow distributions when the two response functions are used (fill squares and fill circles). The top panel and bottom panels shows the results for $b=8$~fm and $b=11$~fm respectively. The curves represent a fit to Eq.~\ref{eq:hia0b}.}
\end{figure}
\begin{figure}[!t]
\centering
\includegraphics[width=1.0\columnwidth]{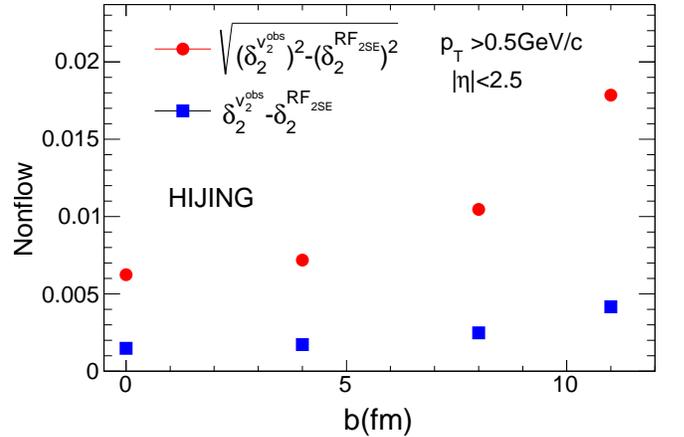}
\caption{\label{fig:hia4} (Color online) Per-particle nonflow estimated for HIJING when the response function RF$_{\mathrm{2SE}}$ is used. The result are obtained by assuming quadrature difference (circles) and linear difference (squares) between the $\delta_n$ values for $v_n^{\;\mathrm{obs}}$ distribution and RF$_{\mathrm{2SE}}$.}
\end{figure}
\begin{figure}[!b]
\centering
\includegraphics[width=1.0\columnwidth]{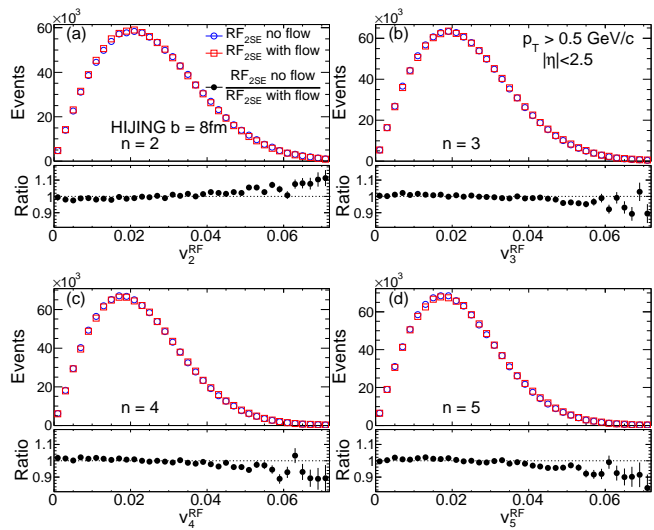}
\caption{\label{fig:hib0} (Color online) The response function using the standard subevents, RF$_{\mathrm{2SE}}$, for HIJING simulations without and with flow for n=2--4. The bottom part of each panel shows the ratio between the two cases.}
\end{figure}

\begin{figure*}[!t]
\centering
\includegraphics[width=0.8\linewidth]{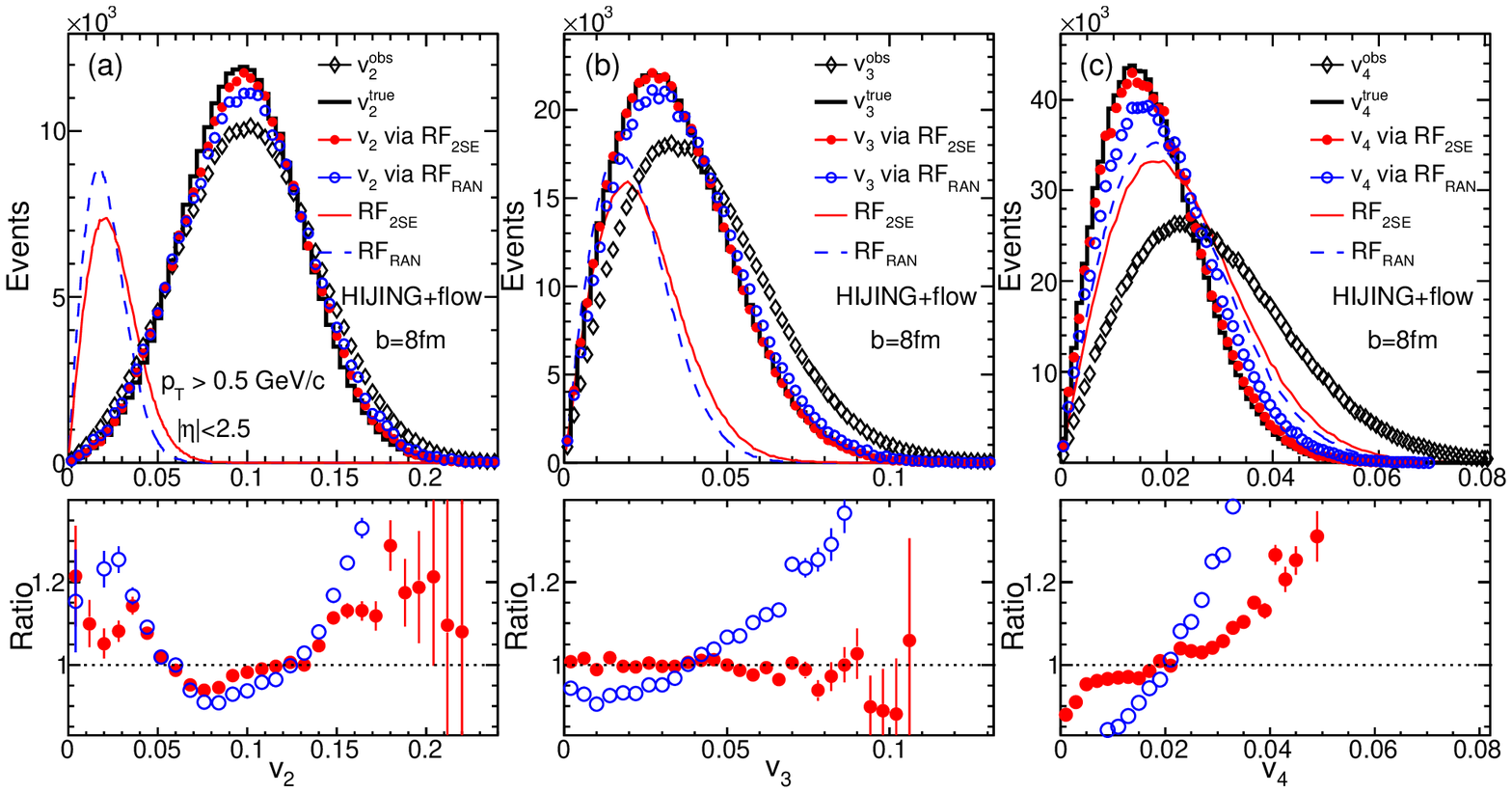}
\caption{\label{fig:hib1} (Color online) The performance of the Bayesian unfolding for $v_2$ (left), $v_3$ (middle) and $v_4$  (right) in $b=8$~fm HIJING events with flow afterburner. The results obtained for RF$_{\mathrm{2SE}}$ and RF$_{\mathrm{RAN}}$ with 128 iterations are shown by the solid squares and open squares, respectively. The bottom panels shows the ratios of the unfolded distributions to the input distributions. Most of the deviations from the input in the tails can be absorbed into a small and simultaneous change of the mean and standard deviation of the $v_n$ distributions (see Figure~\ref{fig:hib4}).}
\end{figure*}
\begin{figure*}[!t]
\centering
\includegraphics[width=0.8\linewidth]{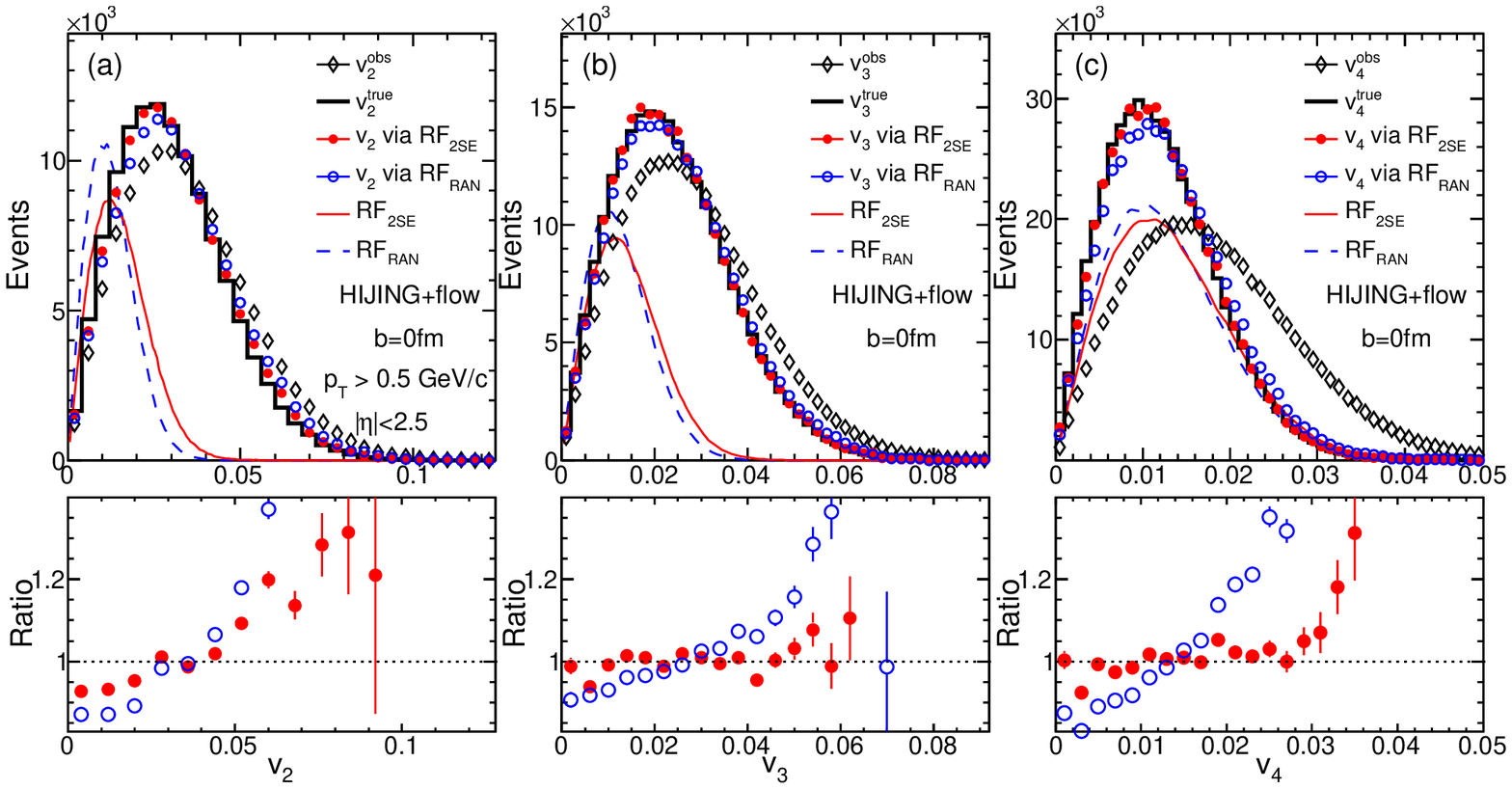}
\caption{\label{fig:hib2} (Color online) The performance of the Bayesian unfolding for $v_2$ (left), $v_3$ (middle) and $v_4$  (right) in $b=0$~fm HIJING events with flow afterburner. The results obtained for RF$_{\mathrm{2SE}}$ and RF$_{\mathrm{RAN}}$ with 128 iterations are shown by the solid squares and open squares, respectively. The bottom panels shows the ratios of the unfolded distributions to the input distributions. Most of the deviations from the input in the tails can be absorbed into a small and simultaneous change of the mean and standard deviation of the $v_n$ distributions (see Figure~\ref{fig:hib4}).}
\end{figure*}
\begin{figure*}[!h]
\centering
\includegraphics[width=0.8\linewidth]{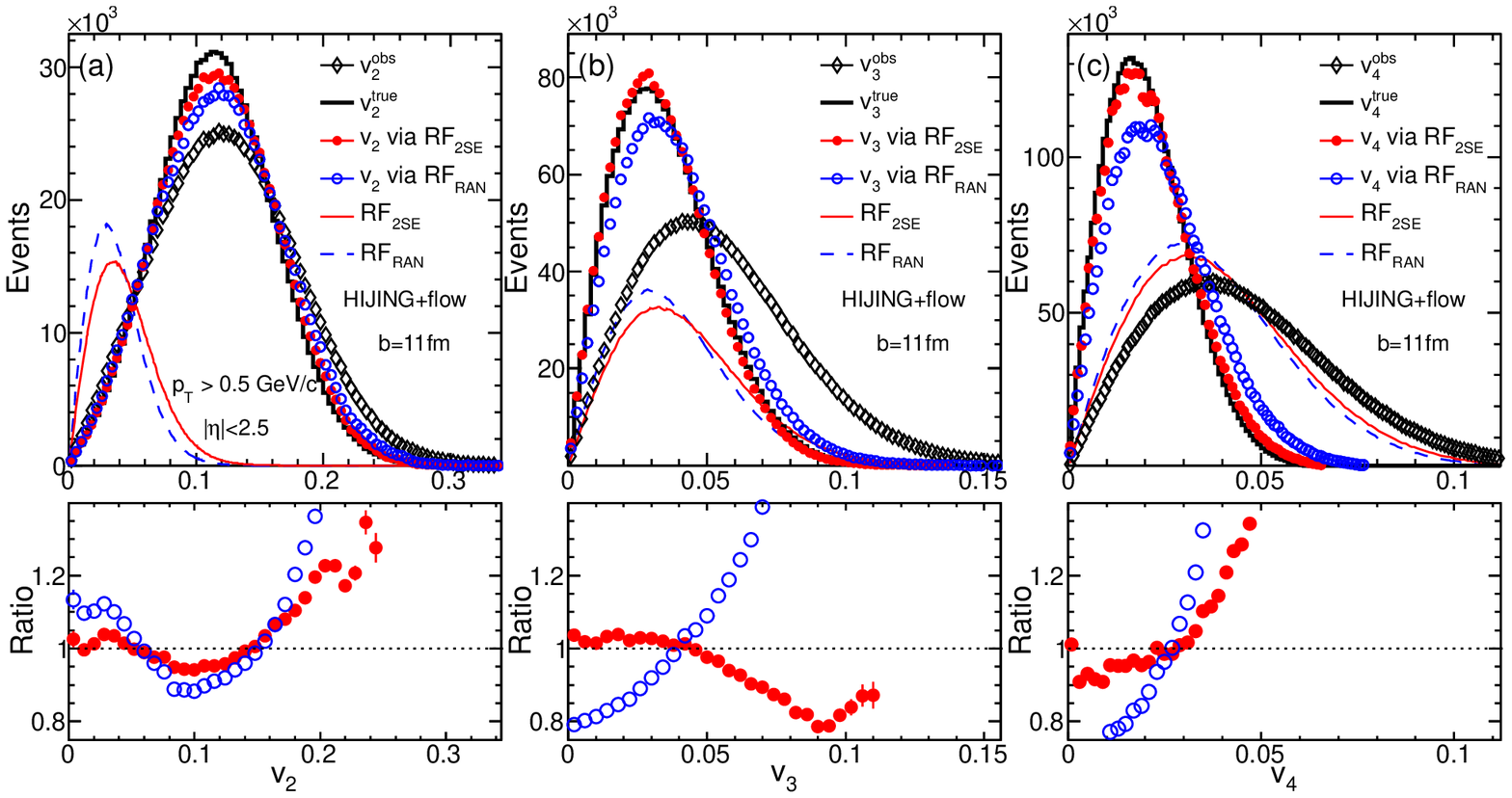}
\caption{\label{fig:hib3} (Color online) The performance of the Bayesian unfolding for $v_2$ (left), $v_3$ (middle) and $v_4$  (right) in $b=11$~fm HIJING events with flow afterburner. The results obtained for RF$_{\mathrm{2SE}}$ and RF$_{\mathrm{RAN}}$ with 128 iterations are shown by the solid squares and open squares, respectively. The bottom panels shows the ratios of the unfolded distributions to the input distributions. Most of the deviations from the input in the tails can be absorbed into a small and simultaneous change of the mean and standard deviation of the $v_n$ distributions (see Figure~\ref{fig:hib4}).}
\end{figure*}
\begin{figure*}[!h]
\centering
\includegraphics[width=1\linewidth]{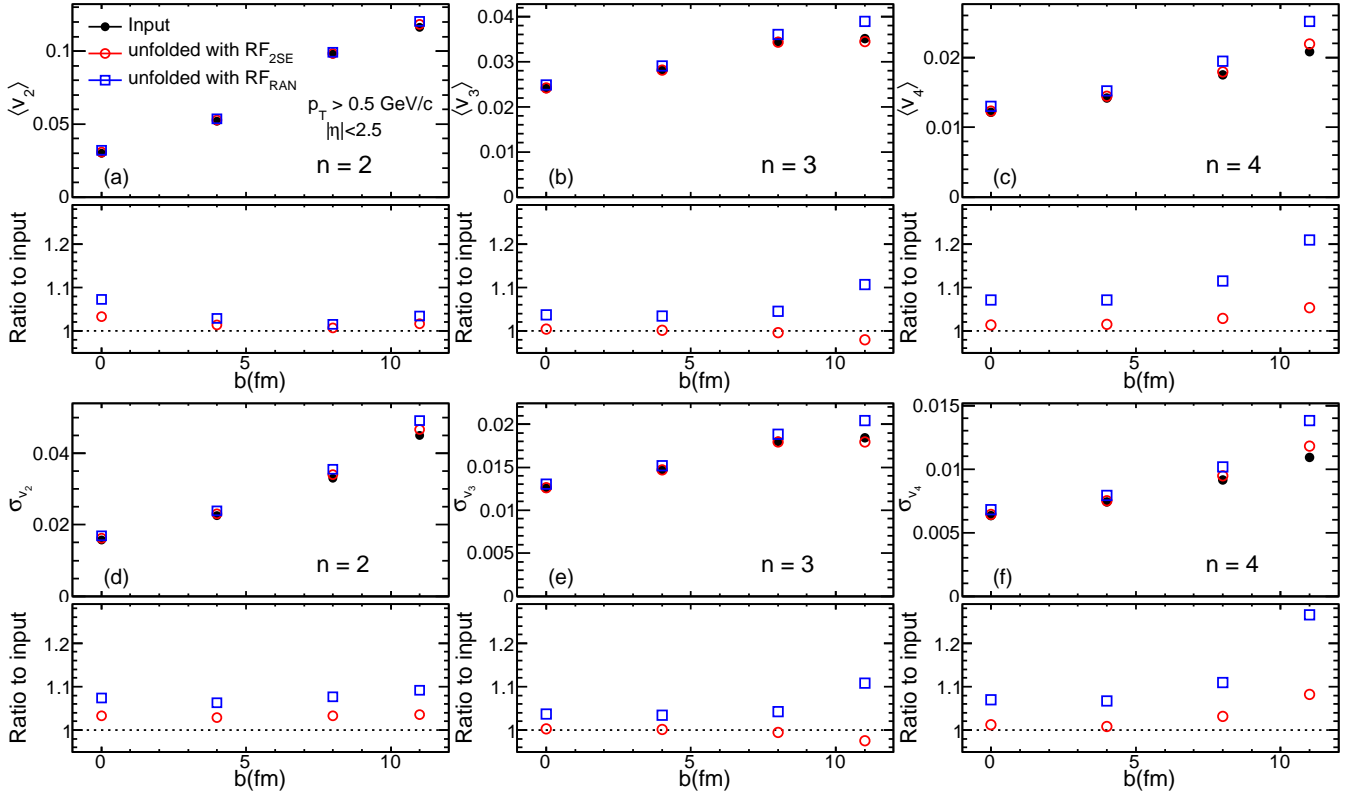}
\caption{\label{fig:hib4} (Color online) The impact parameter dependence of the mean value $\langle v_n\rangle$ and the standard deviation $\sigma_{v_n}$ calculated for the input $v_n$ distribution (solid circles), the $v_n$ distribution unfolded via RF$_{\mathrm{2SE}}$ (open circles) and the $v_n$ distribution unfolded via RF$_{\mathrm{RAN}}$ (open squares). The ratios of these values for the unfolded distribution to those for the input distributions are shown in the bottom part of each panel.}
\end{figure*}
\begin{figure*}[h!]
\centering
\includegraphics[width=1\linewidth]{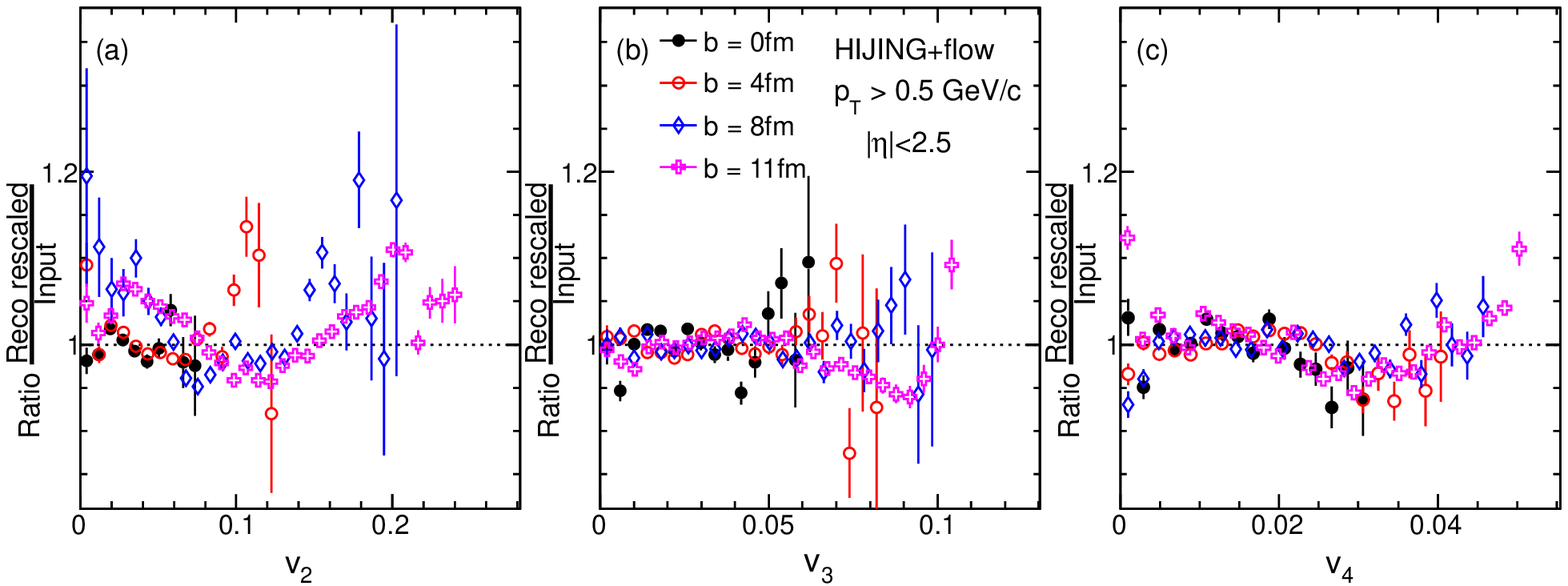}
\caption{\label{fig:hib5} (Color online) The ratios of the unfolded distributions to the input distributions for $v_2$ (left), $v_3$ (middle) and $v_4$ (right) obtained from HIJING+flow simulation for four different impact parameters. The unfolded distributions are obtained for RF$_{\textrm{2SE}}$ (the solid symbols in the top panels of Figures~\ref{fig:hib1}--\ref{fig:hib3}) but have been rescaled horizontally to match the $\langle v_n\rangle$ of the input distributions. The $v_n$ ranges of these ratios are chosen based on the measured $p(v_n)$ distributions in the ATLAS data analysis for comparable centrality intervals, i.e. $b=0$~fm for 0-1\%, $b=4$~fm for 5-10\%, $b=8$~fm for 20-25\% and $b=11$~fm for 50-55\% centrality intervals (see Figure 10 in Ref.~\cite{Jia:2012ve}). The deviations of these ratios from unity are much smaller than those for the ratios obtained without rescaling, as shown by the solid symbols in the bottom panels of Figures~\ref{fig:hib1}--\ref{fig:hib3}.}
\end{figure*}

We found that all distributions in Figure~\ref{fig:hia1} are well described by the radial projection of a 2D Gaussian centered around the origin, i.e. Eq.~\ref{eq:hia0b}. Hence they can be represented by a single parameter $\delta_n$. Figure~\ref{fig:hia2} summarizes the $\delta_n$ versus $n$ for the three distributions for four impact parameters. The widths for RF$_{\mathrm{2SE}}$ are found to agree with that for the $v_n^{\;\mathrm{obs}}$ distributions, except for $n=2$. The small difference seen for $n=4$ may be associated with di-jets that may have a non-zero $v_4$ component. However, all three types of widths are found to approach the statistical limit at large $n$ given by Eq.~\ref{eq:hia1b} (dotted lines).

To quantify the nonflow effects that lead to the difference between the response functions and the $v_n^{\;\mathrm{obs}}$, we unfold the $v_n^{\;\mathrm{obs}}$ distribution using RF$_{\mathrm{2SE}}$ and RF$_{\mathrm{RAN}}$. The unfolded distributions for $n=2$ are shown in Figure~\ref{fig:hia3} for two centrality classes ($b=8$~fm and $b=11$~fm). These distributions can be taken as the EbyE distribution for the remaining nonflow: the distribution obtained with RF$_{\mathrm{RAN}}$ represents the distribution of nonflow beyond pure statistical limit, while the result obtained with RF$_{\mathrm{2SE}}$ represents the distribution of nonflow that are correlated between the subevents (hence not included in RF$_{\mathrm{2SE}}$ and remains after unfolding). The latter is narrower than the former as expected. Both types of nonflow distributions are nearly Gaussian, except in the tails. Similar unfolding exercises are repeated for higher-order harmonics. Nonflow signals are found to be quite small for $n>2$, and their EbyE distributions could not be extracted reliably.

The results shown in Figures~\ref{fig:hia2} and \ref{fig:hia3} allow us to estimate the contributions of nonflow relative to the flow signal, as shown in Figure~\ref{fig:hia4}. They are estimated either as the quadrature difference or as linear difference between the $\delta_n$ values for the $v_n^{\;\mathrm{obs}}$ distribution and RF$_{\mathrm{2SE}}$. The former estimate is appropriate if the nonflow effects contribute as independent smearing on top of the $1/\sqrt{2N}$ smearing. The latter estimate is more appropriate if the nonflow is a overall shift of the $v_n^{\;\mathrm{obs}}$ distribution. The reality probably sits between these two extremes. The worst case example is provided by values for $b=11$~fm (the right most point in Figure~\ref{fig:hia4}): assuming the real flow signal is 0.1, the nonflow contribution is $(0.018/0.1)^2=$3.2\% (quadrature) and 0.004/0.1=4\% (linear). In general, the quadrature (linear) estimation gives a larger nonflow when the flow signal is small (large).

\subsection{HIJING with flow after-burner}
\label{sec:3.2}

To understand how the nonflow effects influence the unfolding performance for extracting the real flow signal, HIJING events are modulated with a $\pT$ dependent flow signal as:
\begin{eqnarray}
\label{eq:hib0}
v_n(\pT) = \frac{\bar{v}_n(\pT)}{\bar{v}_n(p_0)}\; p(v_n)
\end{eqnarray}
where $p(v_n)$ denote the BG function obtained by fitting the ATLAS data as in Figure~\ref{fig:in}, $\bar{v}_n(\pT)$ is the $\pT$-differential $v_n$ measured by the event-plane method~\cite{Aad:2012bu}, and $\bar{v}_n(p_0)$ is the integral $v_n$ for $\pT>0.5$ GeV/$c$ from the event-plane method. The flow signal is implemented by rotating the $\phi$ angle of the particles according to the procedure discussed in Ref.~\cite{Masera:2009zz}. The phases of $v_n$, $\Phi_n$, are randomly distributed, thus there are no correlations between different $n$. The amount of rotation is a smooth function of $\phi-\Psi_n$\cite{Masera:2009zz}, and hence largely preserves the original non-flow correlations. The response functions RF$_{\mathrm{2SE}}$ are unchanged by the implementation of the flow signal as shown in Figure~\ref{fig:hib0}.

The unfolding performance for HIJING with flow is shown in Figure~\ref{fig:hib1} for $b=8$~fm. The unfolding procedure is found to always converge to a stable distribution after a few iterations. But this distribution can be different from the input, as shown in the bottom panels of Figure~\ref{fig:hib1}. The agreement for RF$_{\mathrm{2SE}}$ is systematically better than the result obtained with RF$_{\mathrm{RAN}}$ since some nonflow effects are included in RF$_{\mathrm{2SE}}$. The remaining differences from the input distribution are presumably due to residual nonflow effects not captured by RF$_{\mathrm{2SE}}$. These nonflow effects lead to a broadening of the $v_2$ and $v_4$ distributions, leading to sizable deviations in the tails of the distributions as shown by the ratios in the figure. However, these residual nonflow effects have negligible effects on the $v_3$ distribution. Note that the truth $v_4$ distribution is much narrower than $v_2$ and $v_3$ (see Figure~\ref{fig:in}), hence even a small residual nonflow effect seen in Figure~\ref{fig:hia1} can have a bigger influence on the unfolding performance for n=4 than that for n=2 and 3.

\begin{figure*}[!t]
\centering
\includegraphics[width=1\linewidth]{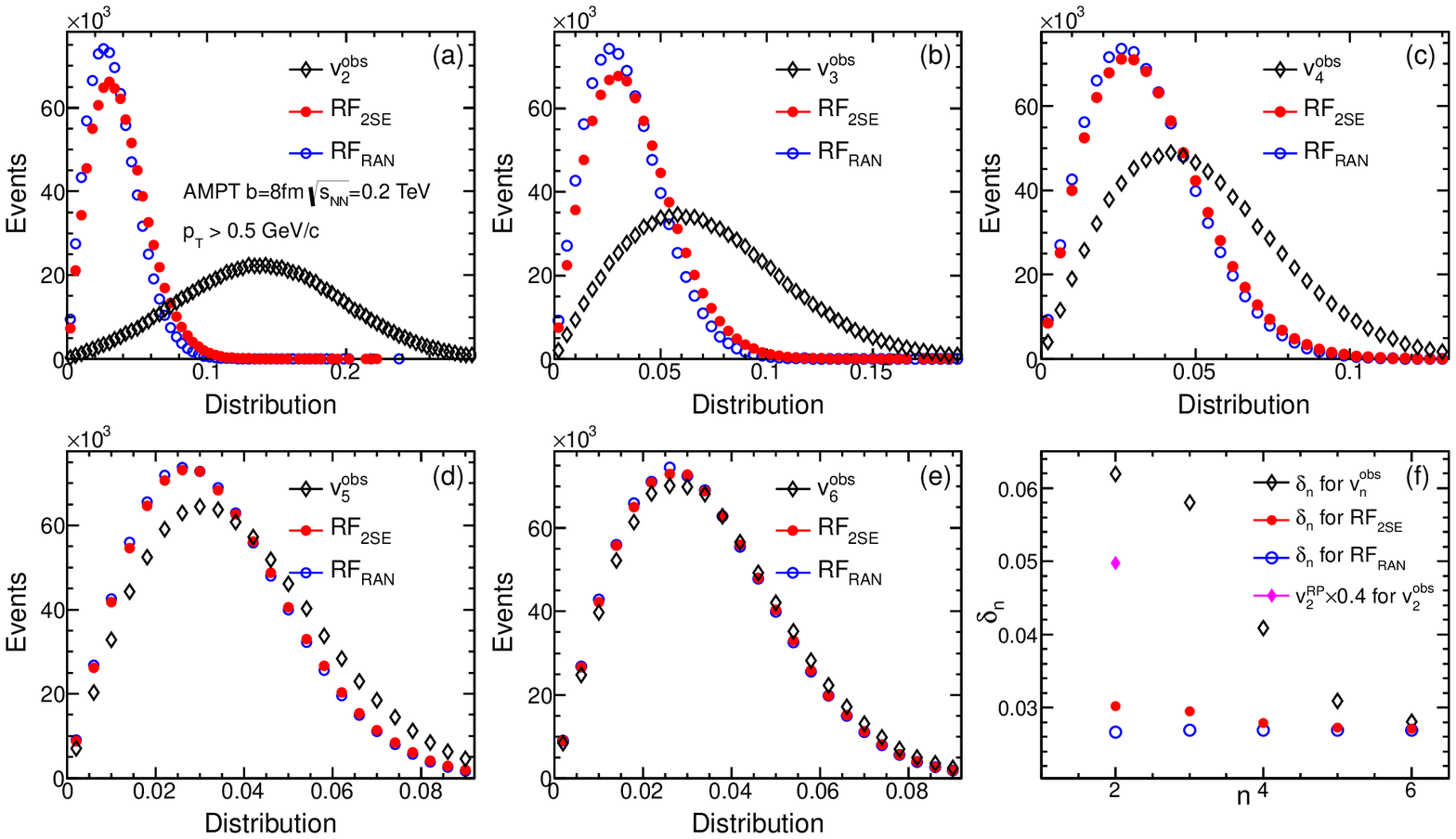}
\caption{\label{fig:hic0} (Color online) The first five panels shows the distributions of $v_n^{\;\mathrm{obs}}$ (diamond), response function obtained via standard 2SE method (squares) and response function obtained via two subevents with randomization of tracks in $\phi$ (circles) for n=2--6. The bottom right panel shows the widths of these distributions vs $n$. The obtained $\delta_2$ and $v_2^{\mathrm{RP}}$ for $v_2^{\mathrm{obs}}$ are obtained by BG fit Eq.~\ref{eq:fluc2}, while all others are fit to Eq.~\ref{eq:hia0b}.}
\end{figure*}

Identical studies are also carried out for other impact parameters. The results for $b=0$~fm and 11~fm are shown in Figure~\ref{fig:hib2} and Figure~\ref{fig:hib3}, respectively. The performance for $b=11$~fm is noticeably poorer, i.e. the final distributions differ significantly from the input $v_n$ distributions in the tails. This is a natural consequence of the influence of the residual nonflow effects not captured in the response functions. 

Figure~\ref{fig:hib4} compares the mean value $\langle v_n\rangle$ and the standard deviation $\sigma_{v_n}$ between the input distribution and the unfolded distributions obtained with RF$_{\mathrm{2SE}}$ and RF$_{\mathrm{RAN}}$. The values from the unfolded distributions are alway larger than those for the input distributions, suggesting that the effects of nonflow always increase the measured $\langle v_n\rangle$ and $\sigma_{v_n}$. The $v_2$ results for RF$_{\mathrm{2SE}}$ agrees with the input within 4\% for all four impact parameters; the agreement for $v_3$ is better than 1\% for $b=0$, 4 and 8~fm, and better than 3\% for $b=11$~fm; the agreement for $v_4$ is better than 2\% for $b=0$ and 4~fm, and worsen to 6\% for $b=11$~fm (consistent with the simple estimation in Figure~\ref{fig:hia4}). 

Most of the deviations from the input values are correlated between $\langle v_n\rangle$ and $\sigma_{v_n}$, which implies that the main difference between the input $v_n$ distribution and the unfolded $v_n$ distribution in Figures~\ref{fig:hib1}-\ref{fig:hib3} can be absorbed into a simultaneous change of $\langle v_n\rangle$ and $\sigma_{v_n}$. Indeed, the shape of the unfolded distribution and the shape of the input distribution, when rescaled to have the same $\langle v_n\rangle$ as shown in Figure~\ref{fig:hib5}, are found to be quite similar. In Figure~\ref{fig:hib5}, the $v_n$ range for the ratios are chosen based on the measured $p(v_n)$ distributions in the ATLAS data analysis in comparable centrality intervals (see Figure 10 in Ref.~\cite{Jia:2012ve}). The deviation from unity in the ratios reaches maximum in the tails of the $v_n$ distributions; They are on the order of 5-15\% depending on the choice of $n$ and the centrality interval, but typically are much smaller than the total systematic uncertainties quoted for the shape of the $p(v_n)$ distributions in the ATLAS data analysis (see Figures 10 and 18--20 in Ref.~\cite{Jia:2012ve}).

Thus far, our discussion has focused mainly on $n\geq2$. In principle, the analysis procedure can also be applied to measure the rapidity-even dipolar flow $v_1$, which is found to be large and dominates over the rapidity-odd directed flow signal at LHC energies~\cite{Aad:2012bu}. However the $v_1$ analysis is sensitive to global momentum conservation effects, and a $\pT$ and $\eta$ dependent weight, \mbox{$w(\pT,\eta) = \pT-\langle\pT^2\rangle(\eta)/\langle\pT\rangle(\eta)$}~\cite{Luzum:2010fb,Jia:2012gu}, needs to be applied to each particle while calculating the response functions. We defer this topic to a future study.
\begin{figure*}[!t]
\centering
\includegraphics[width=0.9\linewidth]{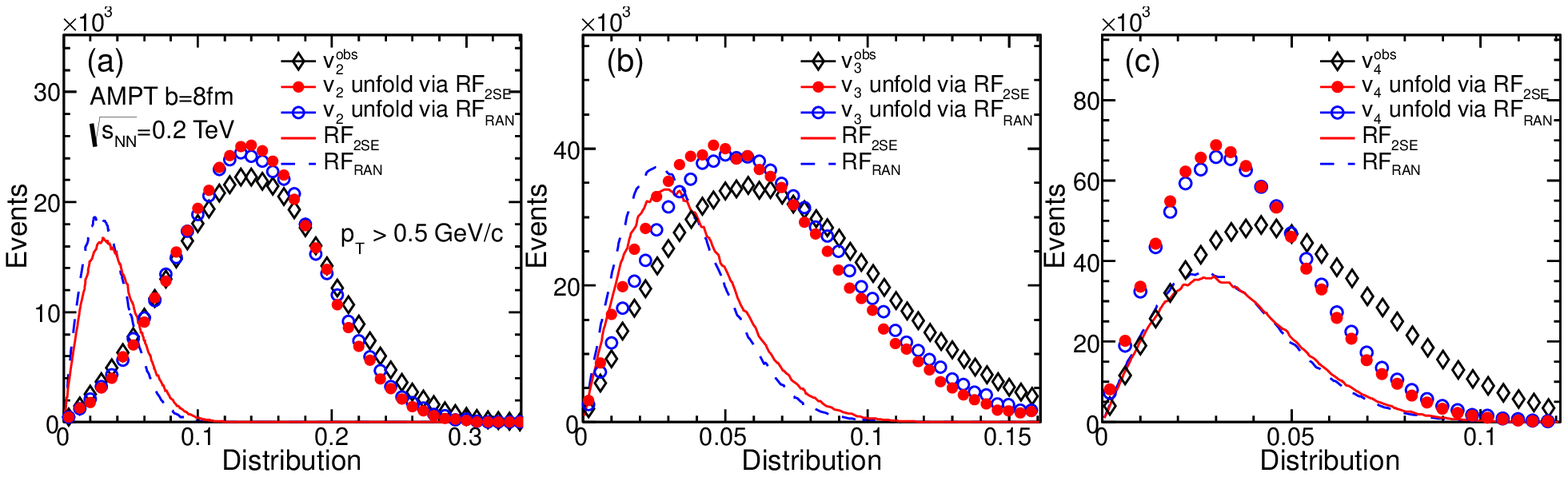}
\caption{\label{fig:hic1} (Color online) The performance of the Bayesian unfolding for $v_2$ (left), $v_3$ (middle) and $v_4$  (right) in $b=8$~fm AMPT events at the RHIC energy. The results obtained for RF$_{\mathrm{2SE}}$ and RF$_{\mathrm{RAN}}$ with 128 iterations are shown by the solid squares and open squares, respectively.}
\end{figure*}
\begin{figure*}[!t]
\centering
\includegraphics[width=0.9\linewidth]{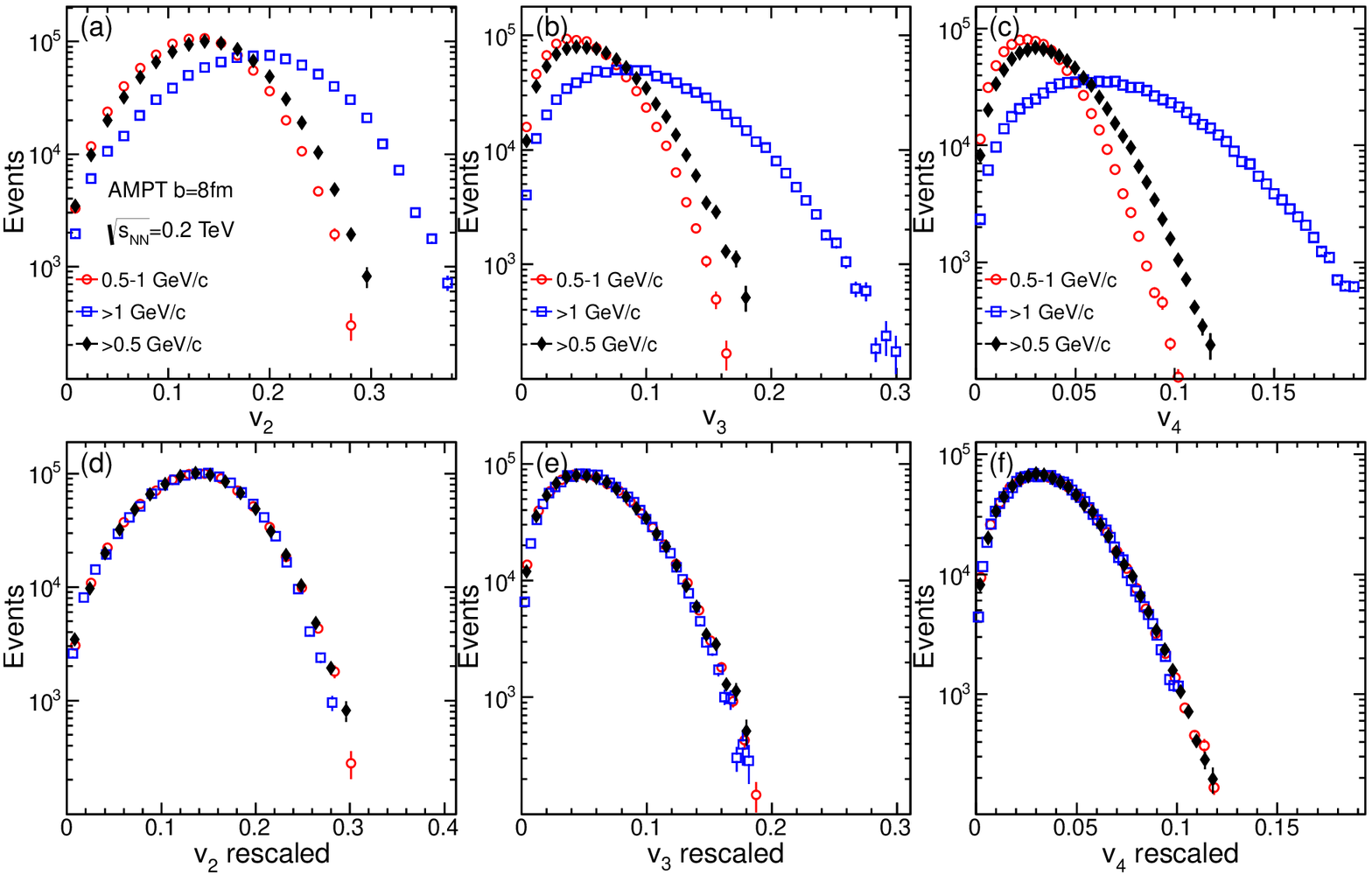}
\caption{\label{fig:hic2}  (Color online) Top panels: The unfolded distributions for $v_n$ in $b=8$~fm AMPT events at RHIC energy for particles in $\pT>0.5$ GeV/$c$, $1>\pT>0.5$ GeV/$c$ and $\pT>1$ GeV/$c$ ranges. Bottom panels: same distributions but rescaled horizontally so the $\langle v_n\rangle$ values match that for $\pT>0.5$ GeV/$c$ range.}
\end{figure*}

\section{AMPT model simulation}
\label{sec:4}
A Muti-Phase Transport model (AMPT)~\cite{Lin:2004en} has been used frequently to study the higher-order $v_n$ associated with $\epsilon_n$ in the initial geometry~\cite{Alver:2010gr,Xu:2011jm,Xu:2011fe,Ma:2010dv}. It combines the initial fluctuating geometry from HIJING and final state interaction via a parton and hadron transport model. The parton and hadron transport is responsible for transforming the $\epsilon_n$ into the momentum anisotropy. The AMPT model naturally contains various nonflow effects present in HIJING, but these nonflow effects can also mix with the collective flow during the transport process. In many aspects, the AMPT model provides a more realistic framework for understanding the role of the nonflow and the hydrodynamic response of the matter to the initial geometry.

The AMPT sample used in this study was generated for $b=8$~fm Au+Au collisions at RHIC energy of $\sqrt{s_{NN}}=0.2$ TeV, containing about 0.8 million events~\footnote{Due to the large amount of resources required for the AMPT simulation, we don't have enough statistics of Pb+Pb collisions at the LHC energy for this study.}. The AMPT events were generated using the parameters that were tuned to reproduce reasonably the experimental $\pT$ spectra and $v_n$ data at RHIC. They were generated with the string-melting mode with a total partonic cross-section of 1.5 mb and strong coupling constant of $\alpha_s=0.33$~\cite{Xu:2011fe}.

Figure~\ref{fig:hic0} shows the distributions of $v_n^{\;\mathrm{obs}}$ and the two types of response functions (RF$_{\mathrm{2SE}}$ and RF$_{\mathrm{RAN}}$) for $n=2$---6. The differences between the two response functions and their dependence on $n$ are quite similar to what is observed in the HIJING simulation (Figure~\ref{fig:hia2}). The large difference in $\delta_n$ between $v_n^{\;\mathrm{obs}}$ and the response functions are due to significant genuine collective $v_n$ signal up to $n=6$. This is clearly revealed in the bottom right panel of Figure~\ref{fig:hic0}. In this case, the $\delta_n$ for $v_n^{\;\mathrm{obs}}$ is obtained via fitting to the BG function:
\begin{eqnarray}
\label{eq:fluc2}
p(x) =\frac{v_n}{\delta_{n}^2}e^{-\frac{(x)^2+(v_n^{\mathrm{RP}})^2}{2\delta_{n}^2}} I_0\left(\frac{v_n^{\mathrm{RP}}x}{\delta_{n}^2}\right)\;,
\end{eqnarray}
$v_n^{\mathrm{RP}}$ is the anisotropy associated with the average shape of the overlap region, while $\delta_{n}$ has contributions from both nonflow and geometry fluctuations. $v_n^{\mathrm{RP}}$ is allowed as a free parameter for $n=2$ but is fixed to be 0 for $n>2$.

The unfolding performance are shown in Figure~\ref{fig:hic1}, the influence of nonflow is reflected by the difference between the unfolded results for the two response functions. The difference is small for $v_2$, but become quite sizable in the tails for $v_3$ and $v_4$.

One important finding in the ATLAS analysis is that the shape of the $v_n$ distributions is nearly independent of the $\pT$ of the particles used (in the low $\pT$ region). This is checked directly in the AMPT simulation. The results in Figure~\ref{fig:hic2} suggest an approximate scaling, although significant deviations are observed for $v_2$ in the tails.
\section{Summary}
\label{sec:5}
The performance of the Bayesian unfolding method for extracting the $v_n$ probability distributions, is evaluated with a toy model simulation, as well as with HIJING and AMPT Monte-Carlo model simulations. This unfolding method has been used previously by the ATLAS Collaboration to obtain the $v_n$ distribution for for $n=2-4$. In the absence of nonflow effects, the unfolding method is demonstrated to converge reliably to the input distributions for $v_2-v_4$ for the typical event multiplicity used in the LHC experiments. The effects of the nonflow are evaluated using HIJING simulation with and without a flow afterburner. The nonflow effects in HIJING are found to be largely statistical in nature, which is consistent with them being associated with independent emitters each containing a finite number of particles. The power spectrum for nonflow are found to decrease rapidly with $n$ and approach a purely statistical limit given by $\sqrt{\langle 1/2N\rangle}$ where $N$ is the number of particles used in the event. The majority of these nonflow effects are included in the data-driven response function and hence are removed in the unfolding procedure. The probability distribution of the residual nonflow, attributed partially to the away-side jets, is extracted for $n=2$ and found to be largely Gaussian. The influence of these residual nonflow effects are found to be significant in the tails of the $v_2$ and $v_4$ distributions and for the peripheral collisions. However this influence can be absorbed in to a simultaneous change, by a few percent, of the mean and the width of the $v_n$ distributions, such that the reduced shape is largely invariant (up to 5-15\% maximum deviations are observed for the reduced shape, depending on $n$ and centrality interval). Thus the data-driven unfolding procedure used by the ATLAS Collaboration is rather robust in removing the nonflow contributions and recovering the true long-range flow correlations.

The unfolding technique is also applied to AMPT transport model simulation which contains both the flow fluctuations and nonflow. The $v_n$ distributions are extracted in two $\pT$ ranges in the low $\pT$ region. When these distributions are rescaled to the same mean values, the adjusted shapes are found to be nearly the same, very similar to the observation of the ATLAS measurement. This finding is also similar to recent event-by-event hydrodynamics model calculations~\cite{Gale:2012rq,Niemi:2012aj}.

\section{EbyE distribution of harmonic coefficient as a tool for studying azimuthal correlations in high-energy collisions}
\label{sec:6}
Although we discussed the calculation of EbyE harmonic coefficients $v_n^{\mathrm{obs}}$ and the Bayesian unfolding in the context of studying flow and nonflow in heavy ion collisions. This method can also be used as a general tool to understand the azimuthal correlations in other high energy experiments such as proton-proton or proton-nucleus collisions, as well as for events selected with particular features, such as events with multiple jets or large missing $E_T$. Such studies are still in progress, nevertheless but we briefly describe the idea as follows.

Each collision produce a distribution of particles in azimuth with a per-particle $v_n^{\mathrm{obs}}$ spectrum calculated via Eq.~\ref{eq:flow}. However this spectrum in a single event can not be used directly as it is dominated by the statistical ``noise'' due to the finite number of particles. This problem can be circumvented by calculating the distribution of $v_n^{\mathrm{obs}}$ from many events. Each event class of interest can have its own set of $v_n^{\mathrm{obs}}$ distributions, one for each $n$. The statistical noise for these distributions can then be estimated from the large $n$ limit (in practice, $n =10$ typically is enough, see Figure~\ref{fig:hia2}):
\begin{eqnarray}
\label{eq:final0}
p_{\infty}(v) \equiv \lim_{n\rightarrow \infty} p(v_n^{\mathrm{obs}}).
\end{eqnarray}
This distribution can also be obtained by randomizing the $\phi$ angle of all tracks (hence destroy all azimuthal correlations). This distribution gives a baseline distribution in the absence of any azimuthal correlation. The distribution of lower-order harmonics is a convolution of this ``noise'' distribution with a probability distribution for genuine correlation $p(s_n)$:
\begin{eqnarray}
\label{eq:final1}
p(v_n^{\mathrm{obs}}) = p_{\infty}(v) \otimes p(s_n).
\end{eqnarray}
where the $s_n$ is a per-particle measure of the correlation strength at angular scale $2\pi/n$. Standard unfolding technique can then be used to solve for $p(s_n)$ using the $p_{\infty}$ as the response function. The distributions of $p(s_n)$ in principle should be sensitive to the nature of the azimuthal correlations of the underlying physics processes in each event class.

If one also wants to distinguish correlations that are long range in $\eta$ from those that are short-range, one should use the response function obtained from the two-subevent method (i.e. squares in Figure~\ref{fig:hia2}) in the unfolding. This procedure could be employed to study the long-range correlations (known as the ``ridge'') recently observed in high multiplicity proton-proton~\cite{Khachatryan:2010gv} and proton-lead~\cite{CMS:2012qk,Abelev:2012cya,Aad:2012gla} collisions at the LHC.

We appreciate valuable comments from R.~Lacey. This research is supported by NSF under award number PHY-1019387.
 % do not change

\begin{thebibliography}{29} % do not change
\bibitem{Alver:2010gr}
  B.~Alver and G.~Roland,
  %``Collision geometry fluctuations and triangular flow in heavy-ion
  %collisions,''
  Phys.\ Rev.\  C {\bf 81}, 054905 (2010)
  [Erratum-ibid.\  C {\bf 82}, 039903 (2010)].
\bibitem{Qiu:2011iv}
  Z.~Qiu and U.~W.~Heinz,
  %``Event-by-event shape and flow fluctuations of relativistic heavy-ion collision fireballs,''
  Phys.\ Rev.\  C {\bf 84}, 024911 (2011).
%, arXiv:1104.0650 [nucl-th].
\bibitem{Voloshin:2008dg} 
  S.~A.~Voloshin, A.~M.~Poskanzer and R.~Snellings, ``Relativistic Heavy Ion Physics,'' Vol.~1, p5, Springer-Verlag (2010). arXiv:0809.2949 [nucl-ex].
  %``Collective phenomena in non-central nuclear collisions,'' 
\bibitem{Teaney:2009qa} D.~A.~Teaney, 
% ``Viscous Hydrodynamics and the Quark Gluon Plasma,'' Editors. R. C. Hwa and X. N. Wang.
``Quark Gluon Plasma 4'', p207,  World Scientific (2010). arXiv:0905.2433 [nucl-th].

\bibitem{Adare:2011tg} 
 A.~Adare {\it et al.}  [PHENIX Collaboration],
  %``Measurements of Higher-Order Flow Harmonics in Au+Au Collisions at $\sqrt{s_{NN}} = 200$ GeV,''
  Phys.\ Rev.\ Lett.\  {\bf 107}, 252301 (2011).
% arXiv:1105.3928 [nucl-ex].
\bibitem{star:2013wf} 
  L. Adamczyk {\it et al.} [STAR Collaboration],
  %``Third Harmonic Flow of Charged Particles in Au+Au Collisions at sqrtsNN = 200 GeV,''
 Phys.\ Rev.\  C {\bf 88}, 014904 (2013).
%  arXiv:1301.2187 [nucl-ex].
  %%CITATION = ARXIV:1301.2187;%%
\bibitem{Aamodt:2011by} 
ALICE Collaboration,
  %``Harmonic decomposition of two-particle angular correlations in Pb--Pb
  %collisions at $\mathbf{\sqrt{s_{\rm NN}} = 2.76}$ TeV,''
 Phys.\ Lett.\ B {\bf 708}, 249 (2012).
% arXiv:1109.2501 [nucl-ex].
\bibitem{CMS:2012wg} 
CMS Collaboration,
  %``Centrality dependence of dihadron correlations and azimuthal anisotropy harmonics in PbPb collisions at sqrt(s[NN]) = 2.76 TeV,''
 Eur.\ Phys.\ J.\ C {\bf 72}, 2012 (2012).
% arXiv:1201.3158 [nucl-ex].
\bibitem{Aad:2012bu} 
 ATLAS Collaboration,
  %``Measurement of the azimuthal anisotropy for charged particle production in sqrt(s_NN) = 2.76 TeV lead-lead collisions with the ATLAS detector,''
Phys.\  Rev.\ C {\bf 86}, 014907 (2012).
% arXiv:1203.3087 [hep-ex].

%\cite{Alver:2007qw}
\bibitem{Alver:2007qw} 
  B.~Alver {\it et al.}  [PHOBOS Collaboration],
  %``Event-by-Event Fluctuations of Azimuthal Particle Anisotropy in Au + Au Collisions at $\sqrt{s_{NN}}= 200$ GeV,''
  Phys.\ Rev.\ Lett.\  {\bf 104}, 142301 (2010).
  %[nucl-ex/0702036].
  %%CITATION = NUCL-EX/0702036;%%
  %63 citations counted in INSPIRE as of 28 Jun 2013

\bibitem{Agakishiev:2011eq} 
  G.~Agakishiev {\it et al.}  [STAR Collaboration],
  %``Energy and system-size dependence of two- and four-particle $v_2$ measurements in heavy-ion collisions at RHIC and their implications on flow fluctuations and nonflow,''
  Phys.\ Rev.\ C {\bf 86}, 014904 (2012).
 % [arXiv:1111.5637 [nucl-ex]].
  %%CITATION = ARXIV:1111.5637;%%
%\cite{Abelev:2012di}
\bibitem{Abelev:2012di} 
  B.~Abelev {\it et al.}  [ALICE Collaboration],
  %``Anisotropic flow of charged hadrons, pions and (anti-)protons measured at high transverse momentum in Pb-Pb collisions at $\sqrt{s_{NN}}$=2.76 TeV,''
  Phys.\ Lett.\ B {\bf 719}, 18 (2013).
%  [arXiv:1205.5761 [nucl-ex]].
  %%CITATION = ARXIV:1205.5761;%%
  %32 citations counted in INSPIRE as of 28 Jun 2013

\bibitem{Jia:2012ve} 
  ATLAS Collaboration,
  %``Measurement of the distributions of event-by-event flow harmonics in Pb-Pb Collisions at $\sqrt{s_{NN}}=2.76$ TeV with the ATLAS detector,''
  ATLAS-CONF-2012-114, arXiv:1209.4232 [nucl-ex], arXiv:1305.2942 [hep-ex].
  %%CITATION = ARXIV:1209.4232;%%
%\cite{Aad:2013xma}
%\bibitem{Aad:2013xma} 
 % G.~Aad {\it et al.}  [ATLAS Collaboration],
  %``Measurement of the distributions of event-by-event flow harmonics in lead--lead collisions at $\sqrt{s_{NN}}$=2.76 TeV with the ATLAS detector at the LHC,''
%  arXiv:1305.2942 [hep-ex].
  %%CITATION = ARXIV:1305.2942;%%
  %1 citations counted in INSPIRE as of 29 Jun 2013

\bibitem{Jia:2012sa} 
  ATLAS Collaboration,
  %``Measurement of Event Plane Correlations in Pb-Pb Collisions at $\sqrt{s_{\mathrm{NN}}}$=2.76 TeV with the ATLAS Detector,''
  ATLAS-CONF-2012-049, arXiv:1208.1427 [nucl-ex].
  %%CITATION = ARXIV:1208.1427;%%
  %``Measurement of reaction plane correlations in Pb-Pb collisions at $\sqrt{s_{\mathrm{NN}}}$=2.76 TeV,''

%\cite{Teaney:2012ke}
\bibitem{Teaney:2012ke} 
 D.~Teaney and L.~Yan ,
  %``Non linearities in the harmonic spectrum of heavy ion collisions with ideal and viscous hydrodynamics,''
  Phys.\ Rev.\ C {\bf 86}, 044908 (2012).
 % [arXiv:1206.1905 [nucl-th]].
  %%CITATION = ARXIV:1206.1905;%%
\bibitem{Qiu:2012uy} 
  Z.~Qiu and U.~Heinz,
  %``Hydrodynamic event-plane correlations in Pb+Pb collisions at $\sqrt{s}=2.76$ATeV,''
  Phys.\ Lett.\ B {\bf 717}, 261 (2012).
  %[arXiv:1208.1200 [nucl-th]].
  %%CITATION = ARXIV:1208.1200;%%
  %15 citations counted in INSPIRE as of 26 Mar 2013
\bibitem{Teaney:2012gu} 
  D.~Teaney and L.~Yan ,
  %``Non-linear flow response and reaction plane correlations,''
Nucl.\ Phys.\  A {\bf 904}, 365 (2013).
%  arXiv:1210.5026 [nucl-th].
\bibitem{Gale:2012rq} 
  C.~Gale, S.~Jeon, B.~Schenke, P.~Tribedy, and R.~Venugopalan,
  %``Event-by-event anisotropic flow in heavy-ion collisions from combined Yang-Mills and viscous fluid dynamics,''
  Phys.\ Rev.\ Lett.\  {\bf 110}, 012302 (2013).
  %[arXiv:1209.6330 [nucl-th]].
%\cite{Niemi:2012aj}
\bibitem{Niemi:2012aj} 
  H.~Niemi, G.~S.~Denicol, H.~Holopainen and P.~Huovinen,
  %``Event-by-event distributions of azimuthal asymmetries in ultrarelativistic heavy-ion collisions,''
Phys.\ Rev.\ {\bf C} 87, 054901 (2013).
  %arXiv:1212.1008 [nucl-th].

\bibitem{Agostini} G.~D'Agostini,  Nucl.\ Instrum.\ Meth.\ A {\bf 362}, 487 (1995).
% A Multidimensional unfolding method based on Bayes’ theorem, 
\bibitem{unfold} T.~Adye,  %``Unfolding algorithms and tests using RooUnfold,''
  arXiv:1105.1160 [physics.data-an].
%Tim Adye, Kerstin Tackmann, and Fergus Wilson. http://hepunx.rl.ac.uk/\~adye/software/unfold/RooUnfold.html.
\bibitem{Gyulassy:1994ew}
  M.~Gyulassy, X.~-N.~Wang,
  %``HIJING 1.0: A Monte Carlo program for parton and particle production in high-energy hadronic and nuclear collisions,''
  Comput.\ Phys.\ Commun.\  {\bf 83}, 307 (1994).

\bibitem{Lin:2004en} 
  Z.~-W.~Lin, C.~M.~Ko, B.~-A.~Li, B.~Zhang and S.~Pal,
  %``A Multi-phase transport model for relativistic heavy ion collisions,''
  Phys.\ Rev.\ C {\bf 72}, 064901 (2005).
 % [nucl-th/0411110].

\bibitem{Ollitrault:1992bk}
  J.~Y.~Ollitrault,
  %``Anisotropy As A Signature Of Transverse Collective Flow,''
  Phys.\ Rev.\  D {\bf 46}, 229 (1992).
\bibitem{Voloshin:2007pc} 
  S.~A.~Voloshin, A.~M.~Poskanzer, A.~Tang and G.~Wang,
  %``Elliptic flow in the Gaussian model of eccentricity fluctuations,''
  Phys.\ Lett.\ B {\bf 659}, 537 (2008).
  %[arXiv:0708.0800 [nucl-th]].
\bibitem{trk} ATLAS Collaboration, ATLAS-CONF-2011-079, http://cdsweb.cern.ch/record/1355702.
\bibitem{Ollitrault:2009ie} 
  J.~-Y.~Ollitrault, A.~M.~Poskanzer and S.~A.~Voloshin,
  %``Effect of flow fluctuations and nonflow on elliptic flow methods,''
  Phys.\ Rev.\ C {\bf 80}, 014904 (2009).
 % [arXiv:0904.2315 [nucl-ex]].
%\cite{Jia:2006sb}
\bibitem{Jia:2006sb} 
  J.~Jia, [PHENIX Collaboration],
  %``Ways to constrain the away side jet in Au + Au collisions in PHENIX,''
  Nucl.\ Phys.\ A {\bf 783}, 501 (2007).
  %[nucl-ex/0609009].
%\cite{Alver:2010rt}
\bibitem{Alver:2010rt} 
  B.~Alver {\it et al.} [PHOBOS Collaboration],
  %``nonflow correlations and elliptic flow fluctuations in gold-gold collisions at $\sqrt{s_{NN}}=200$ GeV,''
  Phys.\ Rev.\ C {\bf 81}, 034915 (2010).
  %[arXiv:1002.0534 [nucl-ex]].
\bibitem{Kikola:2011tu} 
  D.~Kikola, L.~Yi, S.~Esumi, F.~Wang, and W.~Xie,
  %``Nonflow 'factorization' and a novel method to disentangle anisotropic flow and nonflow,''
  Phys.\ Rev.\ C {\bf 86}, 014901 (2012)
 % [arXiv:1110.4809 [nucl-ex]].
  %%CITATION = ARXIV:1110.4809;%%
%\cite{Xu:2012ue}
\bibitem{Xu:2012ue} 
  L.~Xu, L.~Yi, D.~Kikola, J.~Konzer, F.~Wang, and W.~Xie,
  %``Model-independent decomposition of flow and nonflow in relativistic heavy-ion collisions,''
  Phys.\ Rev.\ C {\bf 86}, 024910 (2012).
 % [arXiv:1204.2815 [nucl-ex]].

%\cite{Masera:2009zz}
\bibitem{Masera:2009zz} 
  M.~Masera, G.~Ortona, M.~G.~Poghosyan and F.~Prino,
  %``Anisotropic transverse flow introduction in Monte Carlo generators for heavy ion collisions,''
  Phys.\ Rev.\ C {\bf 79}, 064909 (2009).
\bibitem{Luzum:2010fb}
  M.~Luzum and J.~Y.~Ollitrault,
  %``Directed flow at midrapidity in heavy-ion collisions,''
  Phys.\ Rev.\ Lett.\  {\bf 106}, 102301 (2011).
 % [arXiv:1011.6361 [nucl-ex]].
\bibitem{Jia:2012gu} 
  J.~Jia, S.~K.~Radhakrishnan, and S.~Mohapatra,
  %``A study of the anisotropy associated with dipole asymmetry in heavy ion collisions,''
 J.\  Phys.\  G: Nucl.\  Part.\  Phys.\  {\bf 40}, 105108 (2013).
%  arXiv:1203.3410 [nucl-th].
  %%CITATION = ARXIV:1203.3410;%%

%\cite{Xu:2011jm}
\bibitem{Xu:2011jm} 
  J.~Xu and C.~M.~Ko,
  %``Higher-order anisotropic flows and dihadron correlations in Pb-Pb collisions at $\sqrt{s_{NN}}=2.76$ TeV in a multiphase transport model,''
  Phys.\ Rev.\ C {\bf 84}, 044907 (2011).
  %[arXiv:1108.0717 [nucl-th]].
%\cite{Xu:2011fe}
\bibitem{Xu:2011fe} 
  J.~Xu and C.~M.~Ko,
  %``Triangular flow in heavy ion collisions in a multiphase transport model,''
  Phys.\ Rev.\ C {\bf 84}, 014903 (2011),
  %``Pb-Pb collisions at $\sqrt{s_{NN}}=2.76$ TeV in a multiphase transport model,''
  Phys.\ Rev.\ C {\bf 83}, 034904 (2011).
  %[arXiv:1101.2231 [nucl-th]].
%\cite{Ma:2010dv}
\bibitem{Ma:2010dv} 
  G.~-L.~Ma and X.~-N.~Wang,
  %``Jets, Mach cone, hot spots, ridges, harmonic flow, dihadron and $\gamma$-hadron correlation in high-energy heavy-ion collisions,''
  Phys.\ Rev.\ Lett.\  {\bf 106}, 162301 (2011).
 % [arXiv:1011.5249 [nucl-th]].

%\cite{Khachatryan:2010gv}
\bibitem{Khachatryan:2010gv} 
CMS Collaboration,
  %``Observation of Long-Range Near-Side Angular Correlations in Proton-Proton Collisions at the LHC,''
  JHEP {\bf 1009}, 091 (2010).
 % [arXiv:1009.4122 [hep-ex]].
%\cite{CMS:2012qk}
\bibitem{CMS:2012qk} 
  CMS Collaboration,
  %``Observation of long-range near-side angular correlations in proton-lead collisions at the LHC,''
  Phys.\ Lett.\ B {\bf 718}, 795 (2013).
  %[arXiv:1210.5482 [nucl-ex]].
  %%CITATION = ARXIV:1210.5482;%%
%\cite{Abelev:2012cya}
\bibitem{Abelev:2012cya} 
ALICE Collaboration,
  %``Long-range angular correlations on the near and away side in p-Pb collisions at sqrt(sNN) = 5.02 TeV,''
 Phys.\ Lett.\ B {\bf 719}, 29 (2013).
% arXiv:1212.2001 [nucl-ex].
%\cite{Aad:2012gla}
\bibitem{Aad:2012gla} 
 ATLAS Collaboration,
  %``Observation of Associated Near-side and Away-side Long-range Correlations in $\sqrt{s_{NN}}=5.02$ TeV Proton-lead Collisions with the ATLAS Detector,''
  Phys.\ Rev.\ Lett.\ {\bf 110} 182302 (2012).
 % [arXiv:1212.5198].
  %%CITATION = ARXIV:1212.5198;%%
\end{thebibliography}
\end{document}